\documentclass[prd,nofootinbib,a4paper,showpacs,showkeys,preprintnumbers,twocolumn]{revtex4-1}
\usepackage{graphicx}
\usepackage{subfig}
\usepackage[latin1]{inputenc}
\usepackage{amsmath}
\usepackage{amsfonts}
\usepackage{amssymb}
\usepackage{setspace}
\usepackage{mathtools}
\usepackage{pstricks}
\usepackage{color}
\providecommand{\f}[2]{\frac{{#1}}{{#2}}}

\newcommand{\ba}{\begin{eqnarray}}
\newcommand{\ea}{\end{eqnarray}}

\newcommand{\Mpl}{M_{\rm pl}}

\begin{document}

\title{Quantum corrections to inflation: the importance of RG-running and choosing the optimal RG-scale}

\author{Matti Herranen$^{a,b,c}$}
\email[\,]{matti.h.herranen@jyu.fi}
\author{Andreas Hohenegger$^{d}$}
\email[\,]{andreas.hohenegger@uis.no}
\author{Asgeir Osland$^{d}$}
\email[\,]{asgeir.osland@uis.no}
\author{Anders Tranberg$^{d}$}
\email[\,]{anders.tranberg@uis.no}


\begin{abstract}
We demonstrate the importance of correctly implementing RG-running and choosing the RG-scale when calculating quantum corrections to inflaton dynamics. We show that such corrections are negligible for single-field inflation, in the sense of not altering the viable region in the $n_s-r$ plane, when imposing Planck constraints on $A_s$. Surprisingly, this also applies, in a nontrivial way, for an inflaton coupled to additional spectator degrees of freedom.  The result relies on choosing the renormalisation scale (pseudo-)optimally, thereby avoiding unphysical large logarithmic corrections to the Friedmann equations and large running of the couplings. We find that the viable range of parameters of the potential is altered relative to the classical limit, and we find an upper limit of $g\simeq 10^{-4}$ on the value of the inflaton-spectator portal coupling still allowing for inflation. And an upper limit of $g\simeq 10^{-5}$ for inflation to correctly reproduce the scalar amplitude of fluctuations $A_s$.
\end{abstract}

\maketitle

%%%%%%%%%%%%%%%%%%%%%%%%%%%%
\section{Introduction}
\label{sec:introduction}
%%%%%%%%%%%%%%%%%%%%%%%%%%%%
 
 Observations of the Cosmic Microwave Background indicate a period of accelerated expansion in the early Universe, known as inflation. The predominant theoretical realisation of such an epoch is through the potential energy of a slowly rolling scalar field, the inflaton. The choice of the potential function $V(\phi)$ largely determines the observable spectrum of scalar and tensor metric fluctuations. 
 
 Traditionally, model building and predictions rely on the classical dynamics of a homogeneous field in an expanding background metric, and in many cases the simplification of expressing evolution equations in terms of a set of slow-roll parameters, equations that may then be truncated at some order in powers of these parameters. But ultimately, this homogeneous field must be identified as the quantum expectation value of a quantum field $\hat{\phi}$ (1-point function, mean field, condensate), subject to a complete quantum field theoretical treatment. Then $V(\phi)$ must be identified as the quantum effective potential\footnote{This is not the low-energy effective potential, obtained from integrating out heavier degrees of freedom, but the quantum effective potential from integrating out all fluctuations and is a function of the mean field only.}.
 
 Various choices of $V(\phi)$ in model building must therefore ideally be connected to an underlying quantum theory, restricting the form that such a potential may take. As an example, it is well known that an interacting theory at tree level generates a tower of interactions at loop level, and that the couplings run with the renormalisation scale. In particular, non-minimal coupling to gravity is generated away from the conformal limit, and field self-interactions generically generate logarithmic dependence on the field.  
 
 Quantum corrections are often argued to be "small" in some sense, and can be controlled by powers of the couplings. For single-field inflation, the self-interactions are typically small to provide the necessary flatness of the potential during the inflationary epoch. On the other hand, the renormalisation scale is a priori arbitrary, and it is tempting to choose it to augment the effects of running \cite{Inagaki:2014wva,Inagaki:2015fva}. This misses the point that the renormalisation scale is a parametrisation of a perturbative diagram truncation. A large dependence on this scale is a signal that perturbation theory is unreliable, and so to trust the computation, it is no longer arbitrary, but subject to a criterion of "small dependence". 
 
Quantum corrections to inflation has been considered several times before (see for instance \cite{Odintsov:1993rt,Bilandzic:2007nb,Tranberg:2008ae,Bezrukov:2010jz,Serreau:2011fu,Garbrecht:2011gu,Enqvist:2011jf,George:2012xs,Markkanen:2012rh,George:2013iia,Enqvist:2013eua,Herranen:2013raa,Markkanen:2013nwa,Garbrecht:2014dca,Markkanen:2014poa,Inagaki:2014wva,Herranen:2014cua,Inagaki:2015fva,Herranen:2015aja,Gautier:2015pca} for a broad range of approaches), with some works reporting significant quantum effects, some negligible effect. Because taking all aspects of the calculation into account is quite involved, often focus is put on a few such aspects, while neglecting or ignoring others (typical examples are the inclusion or omission of RG-running, curvature effects, IR effects and resummations, scalar metric fluctuations). Often, a slow-roll treatment is introduced, which requires that the effective potential in the inflation equation of motion is the same as in the Friedmann equation. This is in general not the case (see for instance \cite{Herranen:2013raa}). 
 
 In this work, we consider two models of inflation: one with a single self-interacting field. And one where this single field is coupled to a spectator field. We compute the 1-loop effective potential and 1-loop (or 2-loop) RG-running, include curvature effects, but neglect IR-effects. 
 The size of the quantum contributions, throughout the evolution, depends manifestly on the RG-scale. As we explain, there is no truely optimal choice of this scale for the second model, but our results suggest one that is preferred, by far, over keeping it constant.
 By performing comprehensive parameter scans using a modified version of MultiModeCode \cite{Price:2014xpa}\footnote{Which does not rely on a slow-roll approximation for solving the field dynamics.}, we investigate for both models whether, when using a (pseudo-)optimal, time-dependent RG-scale, quantum corrections may still be important. We find that for a single inflaton field, the smallness of the self-interaction makes the corrections negligible. But, when coupled to other fields, the size of that coupling is a-priori unconstrained (within reason), potentially leading to a significant modification in the basic CMB predictions. However, we will see that while this is true for the individual trajectory, the overall region of $n_s-r$ produced by these inflation models are largely unchanged when including quantum corrections.
 
 As for many other works on this subject, we will also rely on the "semi-classical" approximation to quantum corrections in curved space-time, where the scalar metric fluctuations are ignored. The quantum scalar field(s) evolve in a classical Friedmann-Robertson-Walker (FRW) background. Including metric fluctuations is discussed in some details in \cite{Herranen:2015aja}, where we argue that for large-field inflation as considered in the present work, the semi-classical approach can be expected to be reliable.
 
 The structure of the paper is as follows: In section \ref{sec:single} we introduce the single-field model, slow-roll and quantum corrections as well as the RG-running and show how to choose the RG-scale. In section \ref{sec:Num1} this model is subject to a numerical sampling and computation of the CMB spectrum including quantum corrections. We analyse the magnitude of the corrections from their projection onto the $n_s-r$ plane. In section \ref{sec:2-field} we extend the single-field model with a "portal" coupling to a second scalar in its vacuum, and describe again the quantum corrections, RG-running and one convenient choice of RG-scale. Section \ref{sec:Num2} is then a numerical sampling of the resulting model. We again consider the impact on the observationally allowed region in an $n_s-r$-diagram, and analyse to what extent our convenient choice of RG-scale is optimal. Section \ref{sec:MMC} is a technical description of the numerical procedure and modifications to MultiModeCode, an existing off-the-shelf package underlying our code. We conclude in section \ref{sec:conclusion}.
 
 %%%%%%%%%%%%%%%%%%%%%%%%%%%%
\section{One self-interacting, massive field, non-minimally coupled to gravity}
\label{sec:single}
%%%%%%%%%%%%%%%%%%%%%%%%%%%%

We consider a single-field inflation model, with the action
\begin{align}
S= & \int d^nx\sqrt{-g} \\
&\times\left[\frac{M_{\rm pl}^2}{2}f(\phi)R+
\frac{1}{2}\partial_\mu\phi\,\partial^\mu\phi-\frac{1}{2}m^2\phi^2-\frac{\lambda}{24}\phi^4
\right],\nonumber
\end{align}
and where we choose the specific case
\ba
f(\phi)=1-\frac{\xi \phi^2}{M^2_{\rm pl}}.
\ea
This is the simplest and most common non-minimal coupling between gravity and matter. 

%%%%%%%%%%%%%%%%
\subsection{{Classical slow-roll}}
\label{sec:classical}
%%%%%%%%%%%%%%%

For a non-minimally coupled theory, the usual slow-roll parameters
\ba
\label{eq:sr1}
\epsilon_1 = -\frac{\dot{H}}{H^2},\qquad
\epsilon_2=-\eta = \frac{\ddot{\phi}}{H\dot{\phi}},
\ea
must be supplemented by (see for instance \cite{Liddle:1994dx,Chiba:2008rp,Chiba:2008ia,Morris:2001ad,Torres:1996fr,Noh:2001ia})
\ba
\label{eq:sr3}
\epsilon_3 =-\frac{\xi\phi\dot{\phi}}{HM_{\rm pl}^2f(\phi)},\,
\epsilon_4=\frac{6\xi\left(\xi-\frac{1}{6}\right)\phi \dot{\phi}}{H\left(M_{\rm pl}^2+6\xi\left(\xi-\frac{1}{6}\right)\phi^2\right)}.
\ea
When all of these are $\ll 1$, the Universe is inflating, and the inflaton field is slow-rolling. It then makes sense to write down and expand the equations of motion and observables in powers of $\epsilon_{1,2,3,4}$.
In a flat FRW background, the classical field equation of motion then reads
\ba
\label{eq:clas1}
3H\dot{\phi}\left(1+\frac{1}{3}\epsilon_2\right)=-\left(m^2+\xi R\right)\phi+\frac{1}{6}\lambda\phi^3,
\ea
with the scalar curvature $R=12 H^2 + 6\dot{H}=12 H^2 (1-\epsilon_1/2)$. This follows from the variation of the action with respect to the field $\phi$. The Friedmann equations follow from variation of the action with respect to a general metric $g^{\mu\nu}$, and subsequently specialising to FRW space. We find
\begin{align}
\label{eq:clas2}
&3M^2_{\rm pl}H^2f(\phi)(1+2\epsilon_3)=\nonumber\\
&\quad\quad\quad\quad\quad=\frac{1}{2}\dot{\phi}^2+\frac{1}{2}m^2\phi^2+\frac{\lambda}{24}\phi^4,\\
\label{eq:clas3}
&3M^2_{\rm pl}H^2f(\phi)\left(\frac{2}{3}\epsilon_1-1-\frac{4}{3}\epsilon_3(1+\frac{1}{2}\epsilon_2)\right) = \nonumber\\
&\quad\quad\quad\quad\quad=\frac{1}{2}(1-4\xi)\dot{\phi}^2-\frac{1}{2}m^2\phi^2-\frac{\lambda}{24}\phi^4.
\end{align}
In addition, we define the quantity
\ba
\label{eq:sr5}
\delta = \frac{V''}{H^2}=\frac{m^2+\xi R+\frac{1}{2}\lambda\phi^2}{H^2},
\ea
which, for a 1-field model of inflation, is related to leading order in slow-roll as $\delta\simeq 3(\epsilon_2-\epsilon_1)$. In the slow-roll limit (i.e.~neglecting $\epsilon_{1,2,3,4}$ relative to constant of order 1), the field- and Friedmann equations become
\ba
\label{eq:SRclas1}
3H\dot{\phi}&=&-\left[\left(m^2+12 \xi  H^2\right)\phi+\frac{1}{6}\lambda\phi^3\right],\\
\label{eq:SRclas2}
3M^2_{\rm pl}H^2f(\phi)&=&\frac{1}{2}m^2\phi^2+\frac{\lambda}{24}\phi^4.
\ea
The standard procedure is then to define the end of inflation as $\epsilon_1=1$, back-track the evolution of $\phi$ and $H$ a time corresponding to $N$ e-folds, and compute the basic CMB observables at this "horizon crossing epoch". These are for a non-minimally coupled model given by
\ba
\label{eq:obs1}
A_s&=&\frac{1}{4\pi^2}\frac{H^4}{\dot{\phi}^2},\\
\label{eq:obs2}
n_s-1&=&-4\epsilon_1-2\epsilon_2+2\epsilon_3-2\epsilon_4,\\
\label{eq:obs3}
r&=&16(\epsilon_1+\epsilon_3),
\ea
for the pivot-scale amplitude $A_s$, scalar spectral index $n_s$ and scalar-to-tensor ratio of amplitudes $r$, respectively. Results based on solving (\ref{eq:clas1}, \ref{eq:clas2}, \ref{eq:clas3}) we will refer to as {\it classical}, to distinguish them from the quantum corrected evolution, we will introduce in the following. Similarly, (\ref{eq:SRclas1}, \ref{eq:SRclas2}) will denote {\it classical slow-roll}. The expressions of the central observables, eqs.~(\ref{eq:obs1}, \ref{eq:obs2}, \ref{eq:obs3}) follow from the evolution of the quantum modes of the metric fluctuations. For the 1-loop, 1PI {\it quantum} treatment we perform in the following, they are unaltered as expressions in slow-roll parameters. They may of course take on different values in case the quantum corrected dynamics produces trajectories in $(\phi,\dot{\phi},H,\dot{H})$ different from the classical ones.

%%%%%%%%%%%%%%%%
\subsection{Quantum corrections}
\label{sec:quantum}
%%%%%%%%%%%%%%%

In the semi-classical approach we may straightforwardly compute the 1-loop effective potential in an FRW background (see for instance \cite{Markkanen:2013nwa}), we find in the $\overline{\textrm{MS}}$-scheme
\begin{align}
\label{potential}
V_{\rm eff}=&\frac{1}{2}m^2(\mu)\phi^2(\mu)+\frac{1}{2}\xi(\mu) R\phi^2(\mu)\\
&+\frac{1}{24}\lambda(\mu)\phi^4(\mu)+\frac{1}{64\pi^2}M^4(\phi)\left[
\log\frac{|M^2(\phi)|}{\mu^2}-\frac{3}{2}
\right].\nonumber
\end{align}
We have introduced the mass (squared) $M^2(\phi)$, which appears in the conformal-time field mode equation.\footnote{It is not the second derivative of the potential.} It reads
\ba
M^2(\phi)=m^2(\mu)+\frac{1}{2}\lambda(\mu)\phi^2(\mu)+\left(\xi(\mu)-\frac{1}{6}\right)R.\,
\ea
The couplings $m^2(\mu)$, $\lambda(\mu)$, $\xi(\mu)$ and the field $\phi(\mu)$ are now running with the renormalisation scale $\mu$, relative to some reference scale $\mu_0$. In the following, we will suppress the explicit $\mu$ in our notation, but keeping in mind that whenever quantum corrections are included, all the parameters run with this scale. In the classical approximation they do not. 
By variation of the effective potential, we find the equation of motion for the mean field
\begin{align}
\label{eq:FieldQuant}
&\ddot{\phi}+3H\dot{\phi}+\left(m^2+\xi R\right)\phi+\frac{1}{6}\lambda\phi^3\nonumber\\
&\quad\quad +\frac{\lambda\phi M^2(\phi)}{32\pi^2}\left(\log\frac{|M^2(\phi)|}{\mu^2}-1\right)=0.
\end{align}
This is an equation for the evolution of the renormalised field $\phi$ and is expressed only in terms of RG-improved couplings. 

The Friedmann equations do not follow from variation of the effective action (\ref{potential}), but from variation of the classical action and, again in the semi-classical approach, computing the appropriate quantum expectation values \cite{Herranen:2015aja}. The result is
\ba
\label{eq:FriedmannQuant}
3M^2_{\rm pl}H^2=T_{00}^C+T_{00}^Q,\nonumber\\
a^23M^2_{\rm pl}H^2\left(\frac{2}{3}\epsilon_1-1\right) = T_{ii}^C+T_{ii}^Q,
\ea
where $T_{\mu\nu}^C$ denotes the classical energy-momentum tensor, and the quantum correction is
\begin{align}
\label{eq:QTmunu}
T_{\mu\nu}^Q=-&g_{\mu\nu}\frac{H^4}{64\pi^2}\left(\log\frac{|M^2(\phi)|}{\mu^2}-1\right)\\
\times&\big[\delta^2-4\delta\epsilon_1-2\delta-6\epsilon_1+12\xi(2-\delta+\epsilon_1-\delta\epsilon_1)\big].\nonumber
\end{align}
Since the object $\delta$ is not a-priori a slow-roll parameter, for illustration we have included contributions to leading order in $\epsilon_1$ and all orders in $\delta$. $\phi$ will from now on be taken to be responsible for the CMB fluctuations, in which case it is a light field in the sense that $\delta$ can be considered leading order in slow-roll. Hence in the following, we will discard three of the eight terms in (\ref{eq:QTmunu}) as higher order in slow-roll ($\delta^2$, $-4\delta\epsilon_1$, $12\xi\delta\epsilon_1$).

%%%%%%%%%%%%%%%
\subsection{RG-running and the choice of RG-scale}
\label{sec:RGrunning}
%%%%%%%%%%%%%%%

The RG-improved couplings $\lambda$, $m^2$, $\xi$ follow from solving the 2-loop\footnote{Note that in the third equation for $\xi$, the 2-loop contribution is the sum of terms proportional to $\lambda^2$.} RG equations \cite{Inagaki:2014wva,Inagaki:2015fva}
\ba
\frac{1}{\mu}\frac{d\lambda}{d\mu}&=&\frac{3\lambda^2}{(4\pi)^2}\left(1-\frac{17}{9}\frac{\lambda}{(4\pi)^2}\right),\\
\frac{1}{\mu}\frac{dm^2}{d\mu}&=&m^2\frac{\lambda}{(4\pi)^2}\left(1-\frac{5}{6}\frac{\lambda}{(4\pi)^2}\right),\\
\frac{1}{\mu}\frac{d\xi}{d\mu}&=&\left(\xi-\frac{1}{6}\right)\frac{\lambda}{(4\pi)^2}\left(1-\frac{5}{6}\frac{\lambda}{(4\pi)^2}\right)+\frac{\lambda^2}{18(4\pi)^4}.\nonumber\\
\ea
The solution is not easily written in closed form, and although we include it in our numerical integration, some of the analysis below will be performed at 1-loop for illustration.
The renormalised field $\phi$ is also a function of the scale through the anomalous dimension,
\ba
{\gamma = \f{\lambda^2}{12(4\pi)^4}\,,}
\label{gamma-function}
\ea
by means of wave-function renormalization:
\ba
\phi(\mu)=\phi_c e^{-\int_{\mu_0}^\mu \frac{d\mu' \gamma(\mu')}{\mu'}}\,.
\ea
Since the action and dynamics will be expressed entirely in terms of the renormalised field, the "classical" field $\phi_c$ will not enter explicitly.

It remains to choose the renormalisation scale $\mu$, and the reference scale $\mu_0$. One option is to choose a fixed renormalisation scale, constant in time. This is formally completely valid, but may not provide a good approximation of the effective potential for all times. Since the exact effective potential is independent of the choice of $\mu$, any large dependence on this parameter is a sign that the perturbative truncation is unreliable. The prescription is therefore that one should {\it at each time} choose $\mu$ so that the result depends as little as possible on its exact value. In particular, $\mu$ can be time-, field- and/or Hubble-rate dependent. 

We will do the next-best thing, and define\ba
\label{eq:mudef1}
\mu^2=\frac{|M^2(\phi)|}{e}=\frac{|M^2(\phi,H,\lambda,m^2,\xi)|}{e}.
\ea
This is highly convenient, since the Coleman-Weinberg contributions to both mean field and the entire quantum correction to the Friedman equations simply vanish. The only remaining effect of quantum corrections is that the couplings in (\ref{eq:clas1}, \ref{eq:clas2}, \ref{eq:clas3}), in the expressions for the slow-roll formalism (\ref{eq:sr1}, \ref{eq:sr3}, \ref{eq:sr5}) and the observables (\ref{eq:obs1}, \ref{eq:obs2}, \ref{eq:obs3}) are identical to the classical case, but with couplings that run with the RG-scale. 
The cost is that cancelling the Coleman-Weinberg part, we have committed to a (time-dependent) choice of renormalisation scale, and we may no longer vary it. We note that the fact that the same choice of scale removes quantum corrections from both field equation and Friedmann equation is not generic. We will see an example below where it does not work out, when we consider a 2-field model of inflation.

In order to solve the RG equations, we still need to define the reference scale $\mu_0$, where the parameters have values $\lambda_0,m^2_0,\xi_0$. 
We choose to pick the scale $\mu_0$ to correspond to the initial value of $\phi$, $\phi_0$, deep in the slow-roll inflationary regime. Then using the expression for $\mu$ defined above, we take
\ba
\label{eq:mu02}
\mu_0^2=\mu^2(0)=\frac{|M^2(\phi_0,H_0,\lambda_0,m^2_0,\xi_0)|}{e}.
\ea
This is an explicit expression for $\mu_0$, if we for $\phi_0$ use the slow-roll approximation (\ref{eq:SRclas2}) to determine $H$, $\dot{\phi}$ and $\dot{H}\propto\epsilon_1=0$.

%%%%%%%%%%%
\section{Single-field numerical analysis}
\label{sec:Num1}
%%%%%%%%%%%%%

%%%%%%%%%%%%%%
\subsection{Classical evolution, minimally coupled}
\label{sec:class1}
%%%%%%%%%%%%%

\begin{figure}
\begin{center}
\includegraphics[width=8cm]{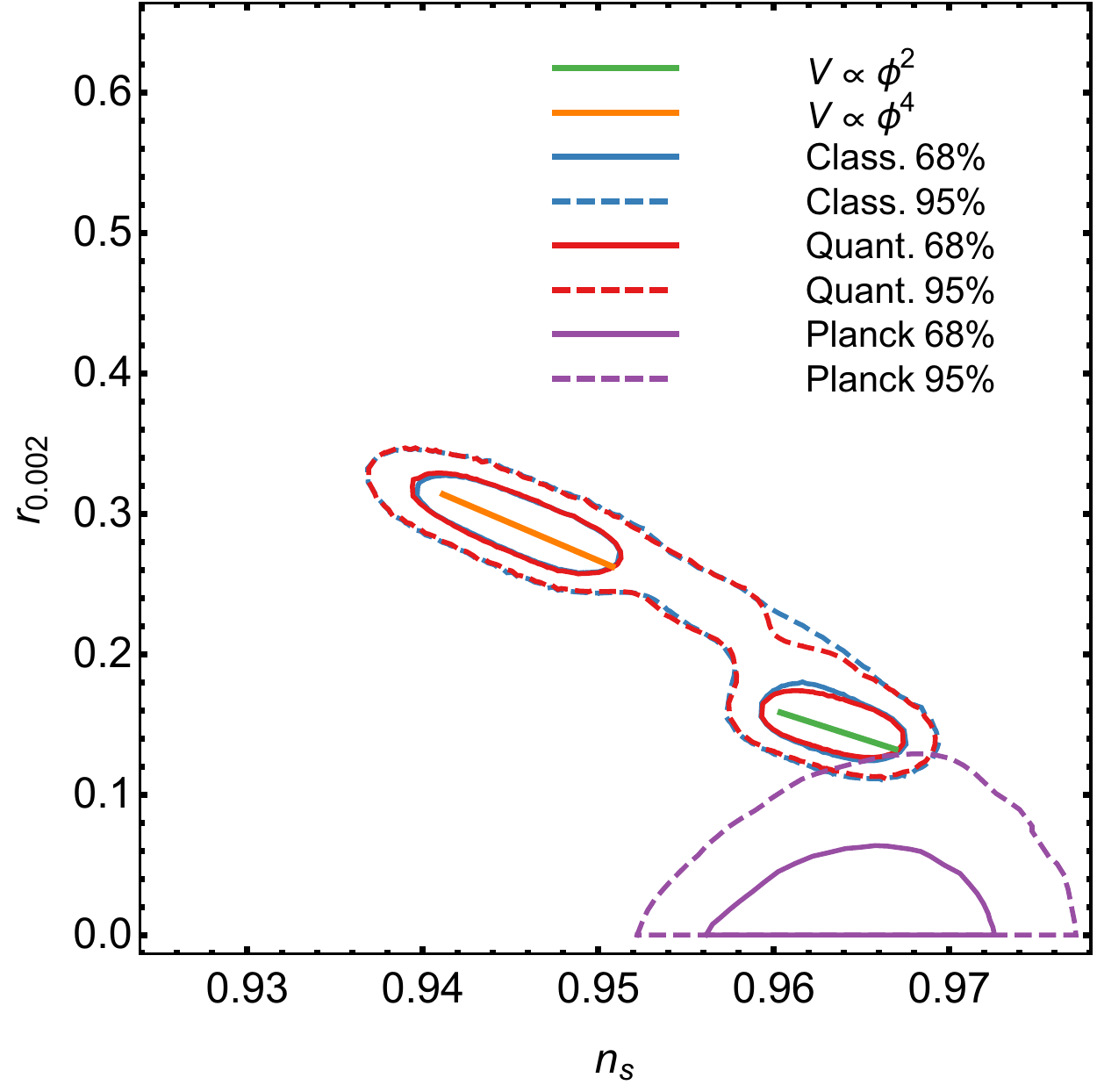}
\caption{The Planck allowed region in the $n_s$-$r$ plane (purple). Overlaid, the slow-roll result for monomial  $\phi^2$ (green) and $\phi^4$ (orange) inflation. Also, non-slow-roll confidence intervals for the classical $\phi^2+\phi^4$ model (blue), and when quantum corrections are taken into account (red). }
\label{fig:1field}
\end{center}
\end{figure}

In the classical limit, to get us started we can with relative ease solve the slow-roll equations semi-analytically, in either of the monomial limits ($m^2=0$, $\xi=0$) or ($\lambda=0$, $\xi=0$). Using the interval $N_*=50$ to $60$ we for each set of parameter values (either $m^2$ or $\lambda$) compute $n_s$ and $r$, imposing the central value of the Planck constraint on $A_s$ \cite{Ade:2015lrj}
\ba
\label{eq:Asranges}
 \log(10^{10} A_s)=3.089 \pm 0.036 \,.
\ea
This is shown as the two line segments in Fig.~\ref{fig:1field}. We confirm the familiar result that classical $\phi^4$ inflation is firmly outside the observationally allowed region (shown as purple contours), and $\phi^2$-inflation marginally so. 

Overlaid are the results of a full numerical sampling, using classical dynamics where we allow to vary both $\lambda$, $m^2$ (so not restricting to monomial inflation) and the initial value $\phi_0$ in the intervals (see section \ref{sec:MMC} for details)
\begin{eqnarray}
\label{eq:ranges}
\lambda_0\in [10^{-16};10^{-10}],\, \frac{m^2_0}{M_{\rm pl}^2}\in [10^{-16};10^{-10}], \, \frac{\phi_0}{M_{\rm pl}}\in [2;30].\nonumber\\
\end{eqnarray}
Note that, although we quote parameters with subscript $0$, in the classical simulation, these do not RG-run and so are equal to $\lambda$, $m^2$ (and shortly also $\xi$ at all scales $\mu$). $\phi_0$ is the initial field value throughout. Also, we do not employ the slow-roll approximation for the field or mode evolution. The slow-roll parameters only enter as we compute the observables to leading order in slow-roll (eqs.~(\ref{eq:obs1}), (\ref{eq:obs2}), (\ref{eq:obs3})). Rather than imposing the central value of $A_s$, we marginalize $A_s$ with a Gaussian distribution of width as in (\ref{eq:Asranges}). These regions are bounded by red lines in Fig.~\ref{fig:1field}.

\begin{figure*}
\begin{center}
\includegraphics[width=\textwidth]{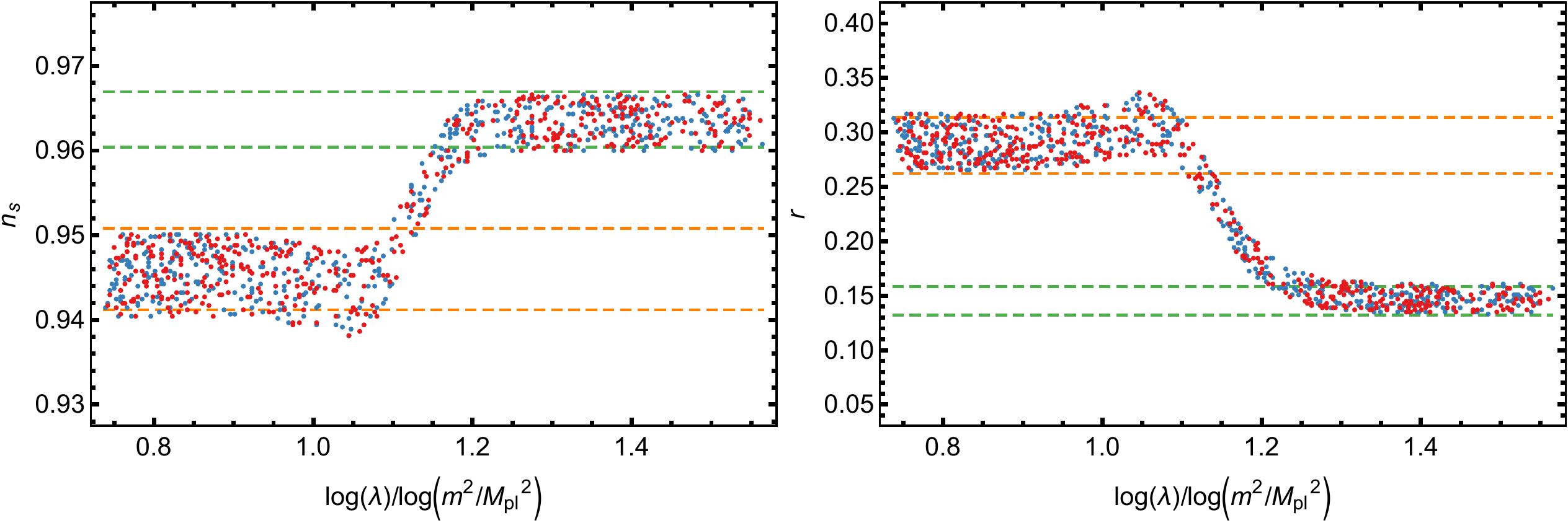}
\caption{Scatter plot showing the dependence of $n_s$ and $r$ on the ratio of the exponents of $\lambda$ and $m^2$. The dominating contribution to the potential typically determines the behaviour of the model as "almost" $\phi^2$ or $\phi^4$ monomial inflation. The horizontal bands indicate the ranges of these models in classical slow-roll for $N_*$ between $50$ and $60$. Red and blue points represent data obtained with and without quantum corrections respectively. Included are only data points whose value for $\log (10^{10} A_s)$ is within nine standard deviations of the Planck experimental value.}
\label{fig:1fieldscatter}
\end{center}
\end{figure*}

We see that the allowed region in the $n_s-r$ is now an elongated "banana", nicely including the semi-analytic SR-intervals inside the 68\% confidence regions. This gives us confidence in the numerical implementation, and also suggests that the slow-roll approximation is in fact rather good. With the given choice of the prior distributions for the parameters, observables close to the classical slow-roll results for monomial inflation are more likely than others. In Fig.~\ref{fig:1fieldscatter}, we see that statistically, either the $\lambda$ or the $m^2$ term dominates the potential and the solution of the field equation behaves accordingly.

%%%%%%%%%%%%%%
\subsection{Quantum evolution, minimally coupled}
\label{sec:quant1}
%%%%%%%%%%%%%

The red contours in Fig.~\ref{fig:1field} refer to a numerical scan of parameters $\xi_0$, $m^2_0$, $\lambda_0$, $\phi_0$ in the exact same ranges as before. But this time, the parameters run as the RG-evolution and scale $\mu$ changes in time as described above. We see that the classical and quantum regions in the $n_s-r$-plane are identical, up to statistical errors. We have checked that there is indeed convergence of the two regions as the statistics increases. In principle, one could imagine having large quantum corrections shifting individual parameter points around in the $n_s-r$ plane, which just happens to create the very same overall distribution. But we have checked that indeed for the individual points, the correction is tiny. 

\begin{figure*}
\begin{center}
\includegraphics[width=\textwidth]{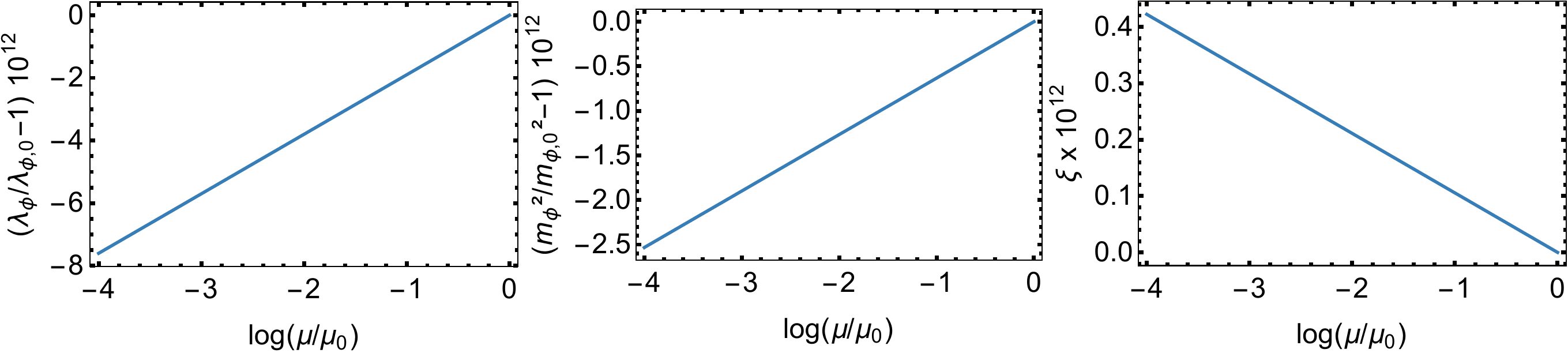}
\caption{The relative 1-loop running couplings for the one-field model. Left: $\lambda/\lambda_0-1$. Middle: $m^2/m_0^2-1$. Right: $\xi$. For all three parameters, we have multiplied by $10^{12}$. For illustration, we have used $\lambda_0=10^{-10}$, $m^2_0=10^{-10} M_{\rm pl}^2$ and $\xi_0=0$.}
\label{fig:RGcouplings1}
\end{center}
\end{figure*}
To further investigate the quantum corrected behaviour, we can analytically solve for the 1-loop running of the couplings as a function of scale $\mu$, finding
\ba
\label{eq:RGsolution1Field1Loop}
\lambda &=& \frac{\lambda_0}{1-\frac{3\lambda_0}{16\pi^2}\log\left[\frac{\mu}{\mu_0}\right]},\nonumber\\
m^2 &=& m^2_0\frac{1}{\left(1-\frac{3\lambda_0}{16\pi^2}\log\left[\frac{\mu}{\mu_0}\right]\right)^{1/3}},\nonumber\\
\xi-\frac{1}{6}&=&\left(\xi_0-\frac{1}{6}\right)\frac{1}{\left(1-\frac{3\lambda_0}{16\pi^2}\log\left[\frac{\mu}{\mu_0}\right]\right)^{1/3}}.
\ea
An example is shown in Fig.~\ref{fig:RGcouplings1}, and we see that for the very small couplings required by observations, the running is very small indeed, unless the scale changes by many orders of magnitude.

To further exemplify the magnitude of the running, let us simply solve for the field evolution in the classical slow-roll approximation, for the case $\xi=0$, $m^2=0$. Let us also, for the sake of argument assume that the initial value of the field is $N_e=60$ e-folds before the end of inflation, the epoch of horizon crossing. Then we have a change of scale during the whole evolution (initial value to end of inflation) of
\begin{align}
\frac{\mu_e}{\mu_0}
=&\left[
\frac{|M^2(\phi_e,H_e)|}{|M^2(\phi_H,H_H)|}
\right]\nonumber\\
=&\left[
\frac{|\frac{1}{2}\lambda \phi_e^2 M_{\rm pl}^2+\frac{1}{12}\lambda(2-\epsilon_1^e)\phi_e^4|}{|\frac{1}{2}\lambda \phi_H^2 M_{\rm pl}^2+\frac{1}{12}\lambda(2-\epsilon_1^H)\phi_H^4|}
\right]\,,
\end{align}
where indices $e$ and $H$ denote the end of inflation and the horizon crossing epoch, respectively, and where we have used the relation
\ba
R= H^2 (12-6\epsilon_1) = \frac{\lambda}{12}\phi^4(2-\epsilon_1).
\ea
Using $\epsilon_1=\phi^2/(8M_{\rm pl}^2)$ for the classical $\lambda\phi^4/24$ model under consideration, we find
\ba
\frac{\mu_e}{\mu_0} = e^{-3.8}\simeq \frac{1}{45}, \qquad \textrm{Monomial $\phi^4$}.
\ea
For $\lambda_0=10^{-10}$, using $\mu=\mu_0/45$, the combination $3\lambda_0\log[\mu/\mu_0]/(16\pi^2)$ featuring prominently in (\ref{eq:RGsolution1Field1Loop}) is $\pm 10^{-11}$, making the running completely negligible. This conclusion does not change upon using 2-loop RG-running, or considering a $m^2\phi^2/2$ theory ($\lambda=0$, $\xi=0$), where 
\ba
\frac{\mu_e}{\mu_0} = e^{-2.4}\simeq \frac{1}{11} , \qquad \textrm{Monomial $\phi^2$}.
\ea
Note that choosing the origin of the running $\mu_0$ to be much deeper in the slow-roll regime (i.e.~much before the horizon crossing epoch) does not matter to the observables $n_s,r$, since they only depend on the value of the potential during and after the horizon crossing epoch. But it does change the mapping from observables to basic variables $\lambda_0,m^2_0$, since they will have run for a while before entering this epoch. Hence, only the running between $\phi_H$ and $\phi_e$ matters. As we see, this is very small, and for the 1-field model, the running is the only quantum effect after choosing the pseudo-optimal RG-scale $\mu(\phi, H, ...)$.

As a result, in a single-field inflation model, quantum corrections are very small. This follows from the running of the couplings, which are all controlled by $\lambda$; and the fact that (as we have seen) any other effect of quantum corrections can be made to disappear by a convenient choice of RG-scale $\mu$. It is tempting to try to compensate the smallness of the coupling by making $\mu$ very small or large, so that the combination $\lambda \log(\mu/\mu_0)$ is much bigger. But that is precisely not allowed, since we have committed to a pseudo-optimal choice of $\mu$ in order to trust our perturbative approximation in the first place. We note that a further reduction of the $\mu$-dependence by finding the truly optimal expression for $\mu$ would only further emphasise this conclusion. We will return to this point shortly. 

We also note that although we are comparing the quantum evolution to a minimally coupled classical evolution, because $\xi$ runs with scale, this is not really well-defined in the quantum case, except if setting the initial value to zero, $\xi_0=0$, as we do here.

%%%%%%%%%%%%%%
\subsection{Non-minimal coupling}
\label{sec:nonminimal}
%%%%%%%%%%%%%

\begin{figure}
\begin{center}
\includegraphics[width=8cm]{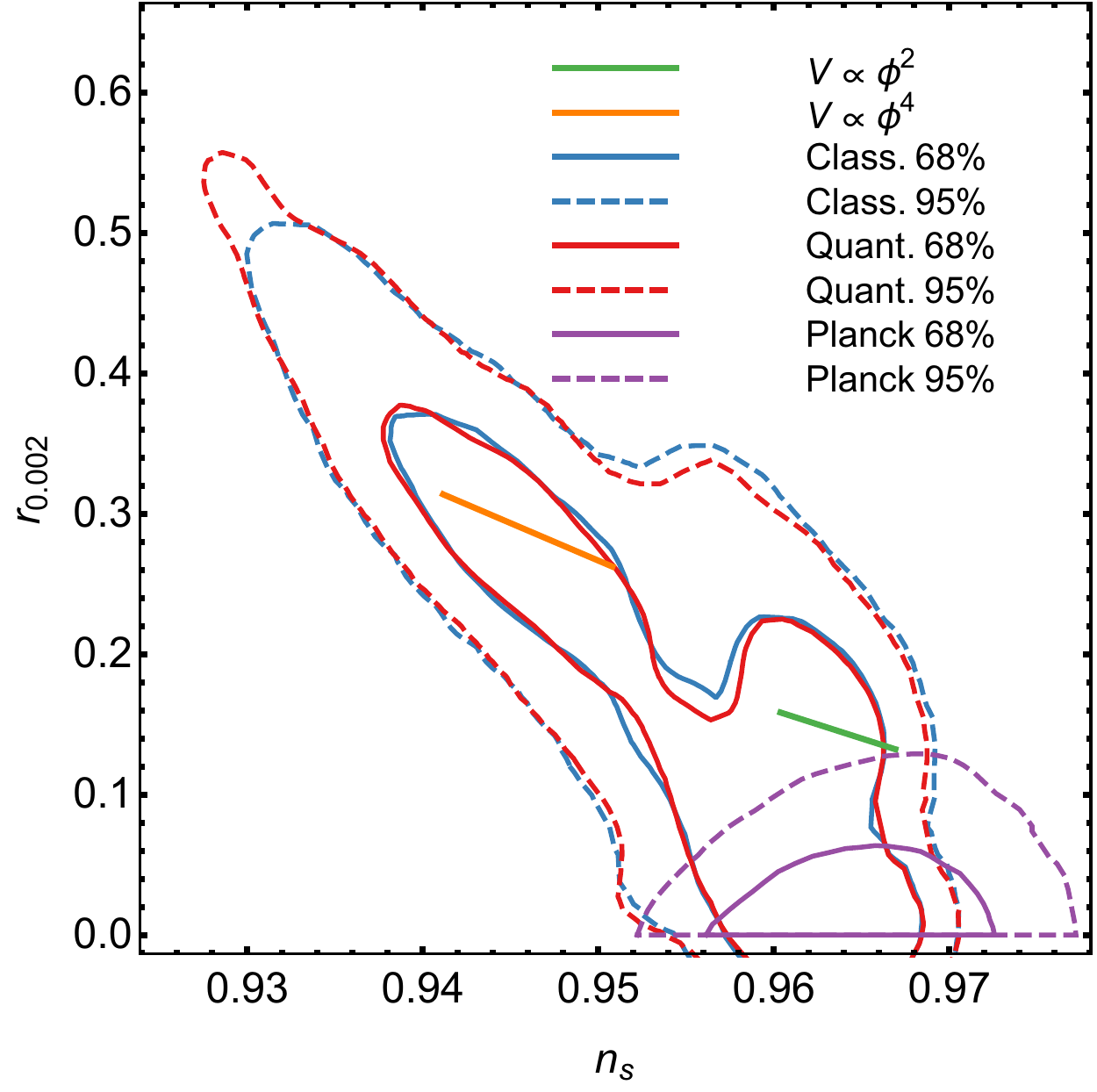}
\caption{The Planck allowed region in the $n_s$--$r$ plane (purple). Overlaid, the slow-roll result for monomial $\phi^2$ (green) and $\phi^4$ (orange) inflation. In addition, confidence intervals for the $\phi^2+\phi^4+\xi R\phi^2$ model using classical dynamics (blue) and quantum corrected dynamics (red).}
\label{fig:1fieldnonmin}
\end{center}
\end{figure}

Allowing now for (initial) non-minimal couplings in the interval 
\begin{eqnarray}
\xi_0\in [10^{-4};1],
\end{eqnarray}
we find Fig.~\ref{fig:1fieldnonmin}. We perform the completely analogous numerical procedure as for the minimally coupled case, running scans with and without quantum corrections.

We again see a familiar result, that allowing for non-minimal coupling, the region consistent with the observed $A_s$ has a substantial overlap with the Planck-allowed region. There is still a weaker correlation between $n_s$ and $r$ along a diagonal in the plot. With our optimized choice of the RG-scale, the quantum corrections do not significantly modify the posterior distribution for the observables. We found that allowing for non-zero initial $\xi$ destroys the simple relationship of Fig.~\ref{fig:1fieldscatter}: Any relative size of $\lambda$ and $m^2$ can be compensated for by a choice of $\xi$ to produce observables inside the Planck-allowed region. The probability distribution in fact takes its largest values close to $r=0$.

%%%%%%%%%%%%%%
\subsection{Validating the choice of $\mu$}
\label{sec:optimal1}
%%%%%%%%%%%%%%%

\begin{figure}
\begin{center}
\includegraphics[width=\columnwidth]{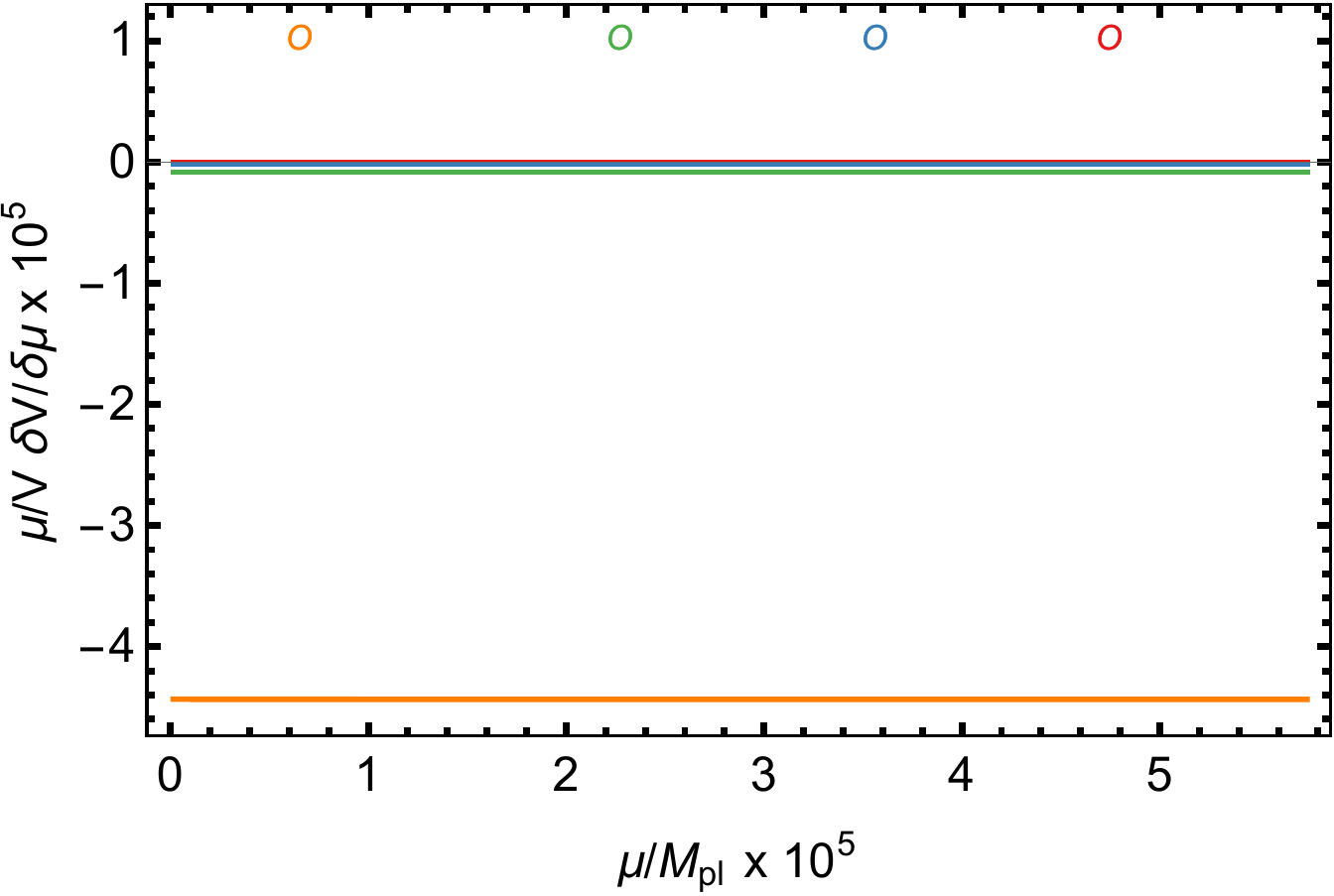}
\caption{The pseudo-optimal scale (circles) at different times (colours), and the relative dependence of the potential on $\mu$ given sets of $(\phi, \dot{\phi}, H, \dot{H})$ along the trajectory. Note the rescaling by $10^5$ along both axes.}
\label{fig:trajmu1}
\end{center}
\end{figure}
Because of the complicated dependence on dynamical variables ($\phi$, $\dot{\phi}$, $H$, $\dot{H}$), one cannot solve for the optimal $\mu$ in closed form
\begin{eqnarray}
\frac{\mu}{V_{\rm eff}}\frac{dV_{\rm eff}}{d\mu}=0.
\label{eq:dVeff}
\end{eqnarray}
It is not even clear, that such a minimum exists. Also, it makes little sense minimising in the complete variables + $\mu$-space, since the dynamical variables are not independent, but follow from a specific inflationary trajectory in field space. 
In Fig.~\ref{fig:trajmu1} we show an example, where we pick 4 times along a given trajectory in $(\phi, \dot{\phi},H, \dot{H})$-space. For each of these, we compute (\ref{eq:dVeff}) as a function of $\mu$ (dashed lines). We see that there is no obvious optimal value, but that the dependence on $\mu$ is very weak at all times during the trajectory. The time-dependent value of the pseudo-optimal scale is indicated as circles. This scale moves towards smaller values during the evolution.

This completes our analysis of the single-field model.

%%%%%%%%%%%%%%%%%%%%%
\section{Coupling to a spectator field}
\label{sec:2-field}
%%%%%%%%%%%%%%%%

For any realistic model of inflation, the inflaton field must be coupled to additional degrees of freedom, and ultimately to the fields of the Standard Model. Because these have quantum fluctuations, their presence is encoded in the effective action of the inflaton itself. But now, because classically the coupling between the two is unconstrained by CMB observables, we can in principle imagine choosing it arbitrarily large, $\mathcal{O}(1)$. There are some constraints due to the reheating mechanism, and how it affects the last few e-folds at the end of inflation. We will not take this complication into account here, since it requires to establish a full field theory simulation. But even for couplings much below unity, our expectation is that the allowed values of the parameters (say, $\lambda_\phi$, $m_\phi^2$, ...) are very much dependent on the inclusion of quantum corrections and the interactions with these other degrees of freedom.

A simple example of this is to consider two scalar fields, the inflaton $\phi$ and a representative of all the other degrees of freedom $\sigma$. Note that $\sigma$ is not intended to be a curvaton. The CMB and the expansion of the Universe both originate from the dynamics of $\phi$. The action reads
\begin{align}
S=\int & d^nx\sqrt{-g}\\
\times\bigg[&\frac{M_{\rm pl}^2}{2}Rf(\phi,\sigma)+
\frac{1}{2}\partial_\mu\phi\,\partial^\mu\phi-\frac{1}{2}m_\phi^2\phi^2-\frac{\lambda_\phi}{24}\phi^4\nonumber\\
&-\frac{g}{4}\phi^2\sigma^2+
\frac{1}{2}\partial_\mu\sigma\,\partial^\mu\sigma-\frac{1}{2}m_\sigma^2\sigma^2-\frac{\lambda_\sigma}{24}\sigma^4
\bigg],\nonumber
\end{align}
with 
\ba
f(\phi,\sigma)=1-\frac{\xi_\phi\phi^2}{M_{\rm pl}^2}-\frac{\xi_\sigma\sigma^2}{M_{\rm pl}^2},
\ea
and where for consistency, we have again allowed for non-minimal coupling to gravity. We will stipulate that $\phi$ provides the dominant energy component through its mean field being displaced from the potential minimum, whereas $\sigma$ will be taken to be in its vacuum around $\sigma=0$. In this case, the {\it classical} 2-field equations of motion are identical to those of the 1-field model in the previous section, and the observables are computed in exactly the same way. 

%%%%%%%%%%
\subsection{Quantum corrections, two fields}
\label{sec:QC_2fields}
%%%%%%%%%%

Computing the 1-loop effective potential, with 1-loop RG running, we have
\ba
V_{\rm eff}&=&\frac{1}{2}m^2_\phi\phi^2+\frac{1}{2}\xi_\phi R\phi^2+\frac{1}{24}\lambda_\phi\phi^4\\
&&+\frac{1}{64\pi^2}M^4_\phi(\phi)\left[\log\frac{|M^2_\phi(\phi)|}{\mu^2}-\frac{3}{2}\right]\nonumber\\
&&+\frac{1}{64\pi^2}M^4_\sigma(\phi)\left[\log\frac{|M^2_\sigma(\phi)|}{\mu^2}-\frac{3}{2}\right],
\ea
where as advertised we have already imposed that\footnote{$\sigma=0$ is a solution also of the quantum evolution, had we written down and solved for the $\sigma$ equation of motion.} $\sigma=0$. But, contrary to the classical limit, at the quantum level the presence of the spectator field $\sigma$ is apparent in the logarithmic corrections to the effective potential.
We now have two effective mode masses
\begin{eqnarray}
M_\phi^2(\phi) &=& m_\phi^2+\frac{\lambda_\phi}{2}\phi^2+(\xi_\phi-1/6)R,\\
M_\sigma^2(\phi) &=& m_\sigma^2+\frac{g}{2}\phi^2+(\xi_\sigma-1/6)R.
\end{eqnarray}
We expect that $\lambda_\phi$ is small to allow for slow-roll inflation. On the other hand, $g$ need not be, and so generically $M_\sigma^2\simeq g\phi^2/2$, much larger than the Hubble rate $H^2$. Hence for the purposes of CMB observables, $\sigma$ is a heavy field and does not contribute to the density perturbations. This is as expected and a useful simplification.
The equation of motion for $\phi$ follows by variation of the effective action,
\ba
\label{eq:TwoFieldQuant}
\ddot{\phi}+3H\dot{\phi}+&&\left(m^2_\phi+\xi_\phi R\right)\phi+\frac{1}{6}\lambda_\phi\phi^3\\&&+\frac{\lambda_\phi\phi M^2_\phi(\phi)}{32\pi^2}\left(\log\frac{|M^2_\phi(\phi)|}{\mu^2}-1\right)\nonumber\\&&+\frac{g\phi M^2_\sigma(\phi)}{32\pi^2}\left(\log\frac{|M^2_\sigma(\phi)|}{\mu^2}-1\right)=0.\nonumber
\ea
We can now choose a pseudo-optimal RG scale $\mu$ in such a way that the two logarithmic terms cancel out. We have
\ba
\label{eq:mu_2field}
&&\frac{\mu^2}{M_{\rm pl}^2} = e^{-1} \\ &&\times \exp\left({\frac{\lambda_\phi M_\phi^2(\phi) \log\left[\frac{|M^2_\phi(\phi)|}{M_{\rm pl}^2}\right]+ gM_\sigma^2\log\left[\frac{|M_\sigma^2(\phi)|}{M_{\rm pl}^2}\right]}{\lambda_\phi M_\phi^2(\phi)+gM_\sigma^2(\phi)}}\right).\nonumber
\ea
Again, this is not the truly optimal choice, but is a very convenient one. This also trivially means that
\begin{align}
\label{ref:neat_relation}
\left(\log\frac{|M^2_\sigma(\phi)|}{\mu^2}-1\right)=-
\frac{\lambda_\phi M^2_\phi(\phi)}{gM_\sigma^2(\phi)}\left(\log\frac{|M^2_\phi(\phi)|}{\mu^2}-1\right).
\end{align}
As for the single-field case, the Friedmann equations do not follow from variation of the effective action, but from a separate computation, varying the classical action with respect to $g^{\mu\nu}$. There are now two fields, with each their contribution (even after setting $\sigma=0$). We have
\ba
\label{eq:TwoFriedmanQuant}
T_{\mu\nu}^Q=-g_{\mu\nu}\frac{H^4}{64\pi^2}\bigg[ && A_\phi \left(\log\frac{|M^2_\phi(\phi)|}{\mu^2}-1\right)\nonumber\\ + && A_\sigma \left(\log\frac{|M^2_\sigma(\phi)|}{\mu^2}-1\right)\bigg],
\ea
with
\ba
A_\phi=\big[&&\delta_\phi^2-4\delta_\phi\epsilon_1-2\delta_\phi-6\epsilon_1\nonumber\\&& +12\xi_\phi(2-\delta_\phi+\epsilon_1-\delta_\phi\epsilon_1)\big],\\
A_\sigma=\big[&&\delta_\sigma^2-4\delta_\sigma\epsilon_1-2\delta_\sigma-6\epsilon_1\nonumber\\&& +12\xi_\sigma(2-\delta_\sigma+\epsilon_1-\delta_\sigma\epsilon_1)\big].
\ea
The $\epsilon_1$ is defined as before, and we again have the slow-roll-like quantities
\ba
\delta_\phi = \frac{m_\phi^2+\frac{\lambda_\phi}{2}\phi^2 + \xi_\phi R}{H^2},\,
\delta_\sigma = \frac{m_\sigma^2+\frac{g}{2}\phi^2 + \xi_\sigma R}{H^2}.\;
\ea
Note however that $\delta_\sigma$ is not small in the slow-roll sense. $\sigma$ is not a light field.
It is clear that the choice of $\mu$, eq.~(\ref{eq:mu_2field}), no longer makes the quantum corrections vanish in the expression for the energy density (as they did for the 1-field case). However, it does allow us to replace one logarithm by another, in accordance with (\ref{ref:neat_relation}),
to get
\ba
T_{\mu\nu}^Q=-g_{\mu\nu}\frac{H^4}{64\pi^2}&&\left(\log\frac{|M^2_\phi(\phi)|}{\mu^2}-1\right)\nonumber\\ \times && \bigg(A_\phi-
A_\sigma
\frac{\lambda_\phi M^2_\phi(\phi)}{gM_\sigma^2(\phi)}
\bigg),
\ea
It is worth looking at the size of this object, in the simplified case of $\xi_\phi=\xi_\sigma=m_\sigma=\lambda_\sigma=0$ (also neglecting, for now, their reappearance due to running). Then, neglecting all SR-sized quantities, we have
\ba
T_{\mu\nu}^Q&\simeq&-g_{\mu\nu}\frac{H^4}{64\pi^2}\left(\log\frac{|M^2_\phi(\phi)|}{\mu^2}-1\right)\\
&&\times\bigg\{
- \frac{g^2\phi^4}{4H^4}+2\frac{g\phi^2}{2H^2}\bigg\}\frac{\lambda_\phi(m_\phi^2+\frac{\lambda_\phi}{2}\phi^2)}{\frac{g\phi^2}{2}}
,\nonumber\\
&\simeq& g_{\mu\nu}\frac{\lambda_\phi g}{64\pi^2}\left(\frac{m_\phi^2}{2}\phi^2+\frac{\lambda_\phi}{4}\phi^4\right)\left(\log\frac{|M^2_\phi(\phi)|}{\mu^2}-1\right).\nonumber
\label{eq:QMcorr}
\ea
This may be compared to the tree-level contributions
\ba
\simeq g_{\mu\nu}\bigg(\frac{m_\phi^2}{2}\phi^2+\frac{\lambda_\phi}{24}\phi^4\bigg),
\ea
which is seen to dominate as long as
\ba
\frac{g\lambda_\phi}{128\pi^2}\left(\log\frac{|M^2_\phi(\phi)|}{\mu^2}-1\right)\ll \quad \frac{1}{2} \quad \textrm{or}\quad \frac{1}{12}.
\ea
This can easily be accommodated. We emphasise that this criterion follows from the choice of RG-scale $\mu$, since without the substitution of (\ref{ref:neat_relation}), the prefactor of the logarithm would have been $(g\phi^2)^2$, which would dominate the tree-level contribution as soon as $g^2\simeq100  \lambda_\phi $. And this in turn would be possible for quite sensible values of $g$. So, although the RG scale does not allow us to cancel out the quantum corrections completely in this case, it does suppress them to a subleading contribution, compared to contributions at tree level. We will confirm this a posteriori below, and it means that we can again identify the leading quantum corrections to be the RG-running of the couplings.

%%%%%%%%%%%%%%%
\subsection{RG-running for two fields}
\label{sec:RGrunning_2fields}
%%%%%%%%%%%%%%%

\begin{figure*}
\begin{center}
\includegraphics[width=\textwidth]{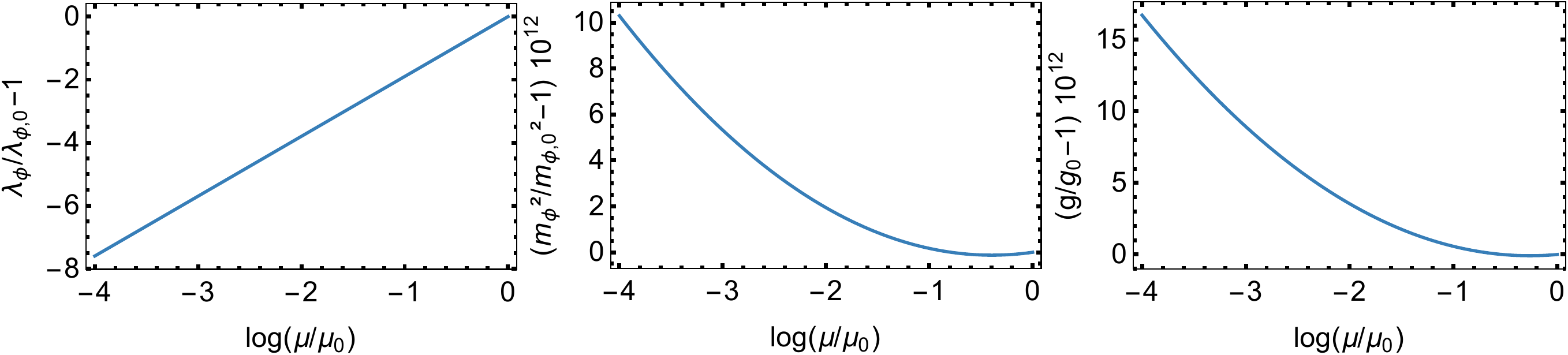}
\caption{The 1-loop running couplings for the two-field model. Whereas $m^2_{\phi}$ and $g$ are again multiplied by $10^{12}$, the relative running of $\lambda_\phi$ is not. For illustration, we have used $\lambda_{\phi,0}=10^{-10}$, $m^2_{\phi,0}=10^{-10} M_{\rm pl}^2$ and $g_0=10^{-4}$. The remaining parameters $\lambda_\sigma$, $m_\sigma^2$, $\xi_{\phi,\sigma}$ also run (not shown).}
\label{fig:2field_RG}
\end{center}
\end{figure*}

There are now seven RG-running couplings $m_\phi^2,m_\sigma^2,\lambda_\phi,\lambda_\sigma,\xi_\phi,\xi_\sigma,g$, and we derive a set of coupled 1-loop RG equations,
\ba
\frac{1}{\mu}\frac{d\lambda_\phi}{d\mu}&=&\frac{3}{(4\pi)^2}\left[\lambda_\phi^2+g^2\right],\nonumber\\
\frac{1}{\mu}\frac{d\lambda_\sigma}{d\mu}&=&\frac{3}{(4\pi)^2}\left[\lambda_\sigma^2+g^2\right],\nonumber\\
\frac{1}{\mu}\frac{dm^2_\phi}{d\mu}&=&\frac{1}{(4\pi)^2}\left[\lambda_\phi m_\phi^2+gm_\sigma^2\right],\nonumber\\
\frac{1}{\mu}\frac{dm^2_\sigma}{d\mu}&=&\frac{1}{(4\pi)^2}\left[\lambda_\sigma m_\sigma^2+gm_\phi^2\right],\nonumber\\
\frac{1}{\mu}\frac{d\xi_\phi}{d\mu}&=&\frac{1}{(4\pi)^2}\left[\lambda_\phi(\xi_\phi-1/6)+g(\xi_\sigma-1/6)\right],\nonumber\\
\frac{1}{\mu}\frac{d\xi_\sigma}{d\mu}&=&\frac{1}{(4\pi)^2}\left[\lambda_\sigma(\xi_\sigma-1/6)+g(\xi_\phi-1/6)\right],\nonumber\\
\frac{1}{\mu}\frac{dg}{d\mu}&=&\frac{1}{(4\pi)^2}g\,\left[\lambda_\phi+\lambda_\sigma\right].
\label{eq:2-field_RG}
\ea
We are mostly interested in the dependence on the coupling $g$, but we see that $\lambda_\phi$, $\lambda_\sigma$ and $g$ source each others RG-running, and so if $g$ is non-zero, they all are. Similarly, only if both masses $m_\phi$ and $m_\sigma$ vanish, do they remain zero. Finally, if both $\xi_\phi$ and $\xi_\sigma$ are equal to their conformal value $1/6$ they do not run. But if either is non-conformal (for instance zero), they both run away from their initial value.

The central observation is that even if $\lambda_\phi$ starts out with a value of $\simeq 10^{(-10, -11,-12)}$, in accordance with the classical observational constraints, it grows semi-linearly as $3/(4\pi)^2 g^2$ $\times\log[\mu/\mu_0]$. Hence even for a reasonable, perturbative value of $g=10^{-4}$, $\lambda_\phi$ will have grown to $>10^{-10}$ with a scale-change of $\log[\mu/\mu_0]= \mathcal{O}(1)$. This would naively suggest that the system is driven out of slow-roll. It is therefore highly conceivable that the inclusion of quantum fluctuations puts strong constraints on the viable values of $g$. And at the same time, a careful choice of $g$ may allow otherwise ruled-out scenarios (such as quartic inflation) to fit observations. We will investigate this further below. In Fig.~\ref{fig:2field_RG} we show one example of the running couplings, using $m^2_{\phi,0}=10^{-10}\,\Mpl^2$, $\lambda_{\phi,0}=10^{-10}$ and $g_0=10^{-4}$.

We observe that all parameters become non-zero. The left panel shows the relative change of the inflaton self-coupling. We see that it can grow by a factor $1-10$ within the expected change of scale $\log[\mu/\mu_0]=-4$. In this example, we have used the maximum value of this initial coupling and the largest value of $g_0$, suggesting that quantum corrections may become important, but need not be. We will see shortly what the effect is. 

We note in passing that the non-minimal coupling remains very small, and that the relative change in the inflaton mass parameter $m^2_\phi$ is very small. Hence, the primary effect of the RG running is for the inflaton self-coupling to change significantly during the time evolution. The inflaton self-coupling $\lambda_\sigma$ grows as large as $\lambda_\phi$, but does not enter in the expressions for the observables.

%%%%%%%%%%%
\section{Two-field numerical analysis}
\label{sec:Num2}
%%%%%%%%%%%%%

%
%%%%%%%%%%
\subsection{Individual trajectories}
\label{sec:indiv2}
%%%%%%%%%%

To illustrate the impact of quantum effects on the cosmological evolution we first choose 3 examples for which inflation is successful. The initial parameters at $\mu=\mu_0$ for these cases are
\ba
\text{BM 1:}\quad&& m_{\phi ,0}^2=6.2\cdot 10^{-11}\,\Mpl^2 ,\, \lambda _{\phi ,0}=5.9\cdot 10^{-14} ,\, \nonumber\\ && g_0=6.4\cdot 10^{-6}\,, \phi_0 =25\Mpl\,,  \nonumber\\
\text{BM 2:}\quad&& m_{\phi ,0}^2=1.5\cdot 10^{-12}\,\Mpl^2 ,\, \lambda _{\phi ,0}=2.7\cdot 10^{-16} ,\,  \nonumber\\ && g_0=2.3\cdot 10^{-10}\,, \phi_0 =21\Mpl\,,  \nonumber\\
\text{BM 3:}\quad&& m_{\phi ,0}^2=1.5\cdot 10^{-13}\,\Mpl^2 ,\, \lambda _{\phi ,0}=4.6\cdot 10^{-15} ,\,  \nonumber\\ && g_0=2.2\cdot 10^{-7}\,, \phi_0 =30\Mpl\,, 
\ea
and $\xi _{\phi ,0} = \lambda _{\sigma ,0} = \xi _{\sigma ,0}= 0$ for BM 1 - BM 3. We choose $N_* =55$ as the pivot scale everywhere and $\epsilon_1 = 1$ as the condition for the end of inflation.

We implement the numerics as for the one-field model. The dynamical variables are the same, but the evolution equations and solving for the time-dependent $\mu$ is more involved. The classical limit is identical to the one-field case (since we are postulating that the second field does not contribute to the energy density of the Universe, nor to the density perturbations). 

For the 3 parameter sets we obtain the following numerical values for the observables:
\ba
\text{BM 1:}&A_s=3.0\cdot 10^{-9} \,, &\delta A_s=-0.046. \nonumber\\
&n_s=0.96 ,\, &\delta n_s=0.0013.\nonumber\\
&r=0.14 ,\, &\delta r =-0.052 \,, \nonumber\\
\text{BM 2:}& A_s=7.7\cdot 10^{-11} \,, &\delta A_s=-3.1\cdot 10^{-9}.\nonumber\\
&n_s=0.96 ,\, &\delta n_s =5.6\cdot 10^{-11}. \nonumber\\
&r=0.15 ,\, &\delta r =1.0\cdot 10^{-9}.\nonumber\\
\text{BM 3:}& A_s=9.3\cdot 10^{-12} \,, & \delta A_s=-0.065 .\nonumber\\
&n_s=0.96 ,\,&\delta n_s =0.00059.\nonumber\\
& r=0.24, &\delta r =0.0056. \nonumber
\ea
$\delta A_s$, $\delta n_s$ and $\delta r$ represent the relative changes compared to the classical results.
We see that the largest differences in these examples are at the percent level.

\begin{figure*}
\begin{center}
\includegraphics[width=0.7\textwidth]{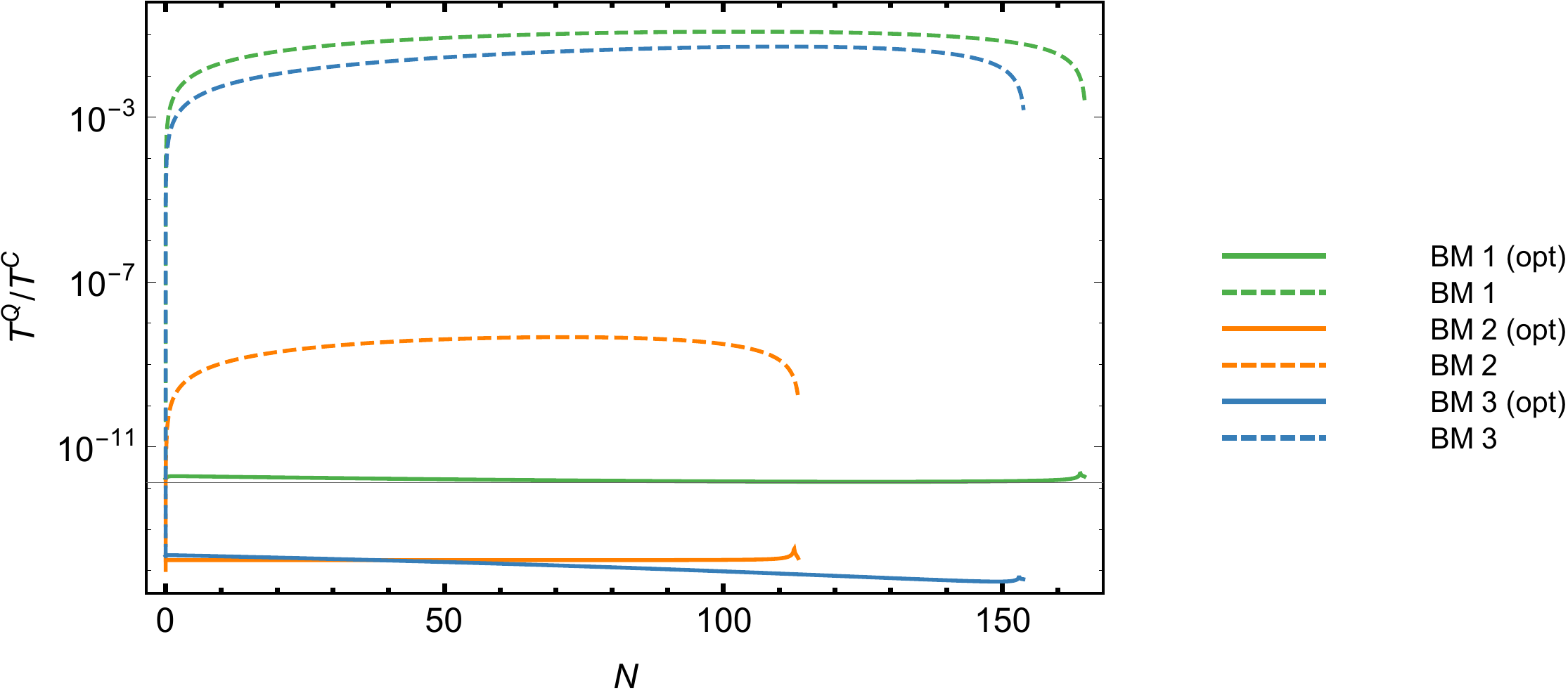}
\caption{Relative size of the quantum corrections to the Friedman equations as a function of the evolved number of e-folds. The dashed lines correspond to these contributions with constant RG-scale, while the full lines show their size if the RG-scale is updated (in both cases computed along the trajectory obtained with only RG-running corrections). }
\label{fig:TwoFieldsQuantumCorrectionsFriedmannN}
\end{center}
\end{figure*}

As described, we update the renormalisation scale to remove the quantum contributions to the field-equation. We find a posteriori that the remnant corrections to the Friedmann equations are negligible with this choice of $\mu$ (see Fig.~\ref{fig:TwoFieldsQuantumCorrectionsFriedmannN}). We show the magnitude along BM1-3 of the logarithmic corrections to the Friedmann equations, relative to the non-logarithmic contributions. When the RG-scale is not adjusted over time as (\ref{eq:mu_2field}) (dashed lines) the contributions are small, but may still be a few percent or more. By adjusting the RG-scale (full lines), we can reduce this to one part in $10^{-11}$ or less. And so although we are not able to identically cancel out the logarithms as for the 1-field model, choosing the RG-scale wisely is a vast improvement on even the "best" constant RG-scale. We therefore ignore these contributions keeping only the RG-running of the couplings according to equations (\ref{eq:2-field_RG}).

\begin{figure*}
\begin{center}
\includegraphics[width=0.7\textwidth]{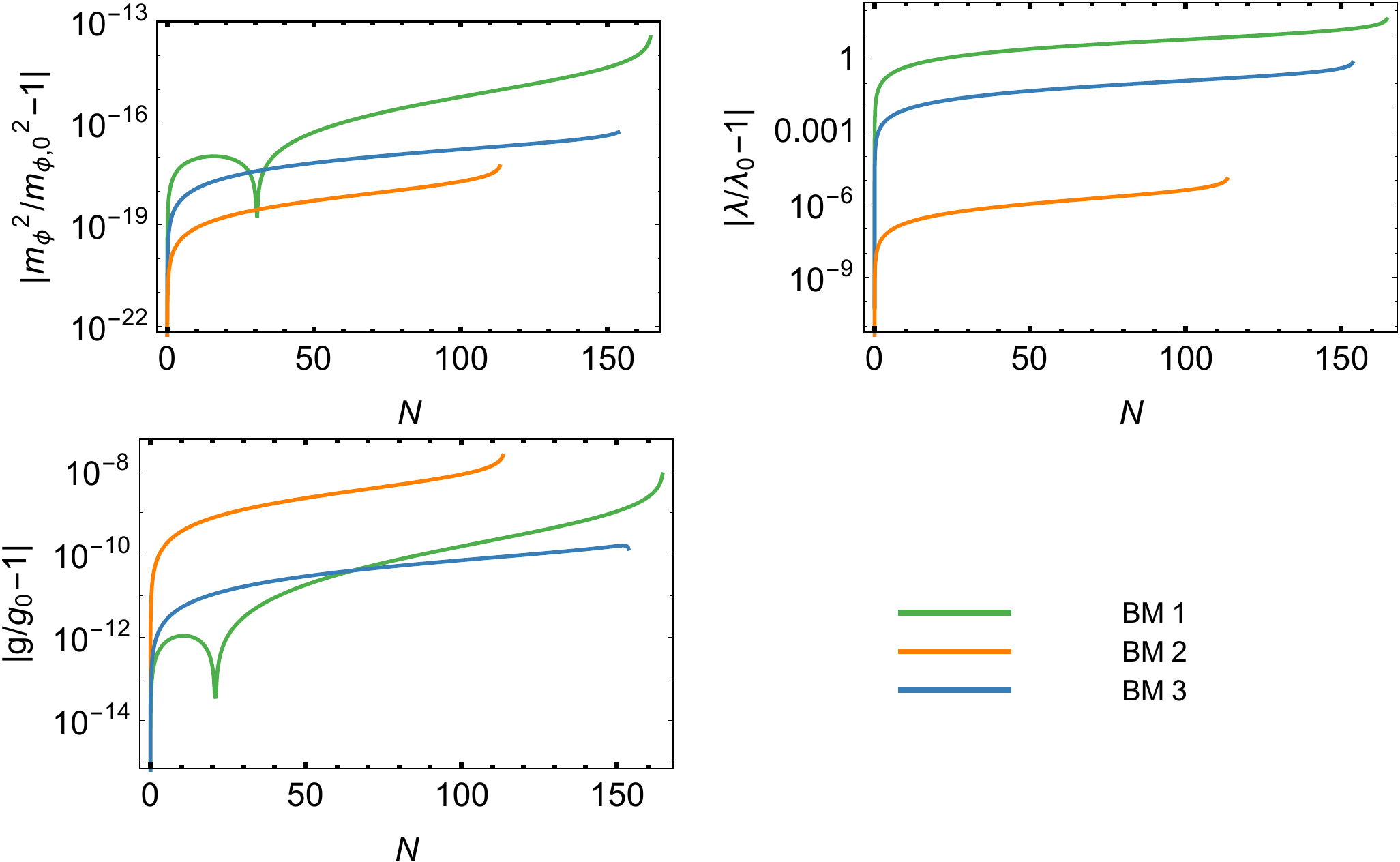}
\caption{Relative change of some of the two-field model parameters as a function of the evolved number of e-folds $N$ for different initial parameters (BM 1 - BM 3). Further parameters that have zero initial values are not shown but run as well. }
\label{fig:TwoFieldsRGrunningBenchmark1}
\end{center}
\end{figure*}

\begin{figure*}
\begin{center}
\includegraphics[width=0.7\textwidth]{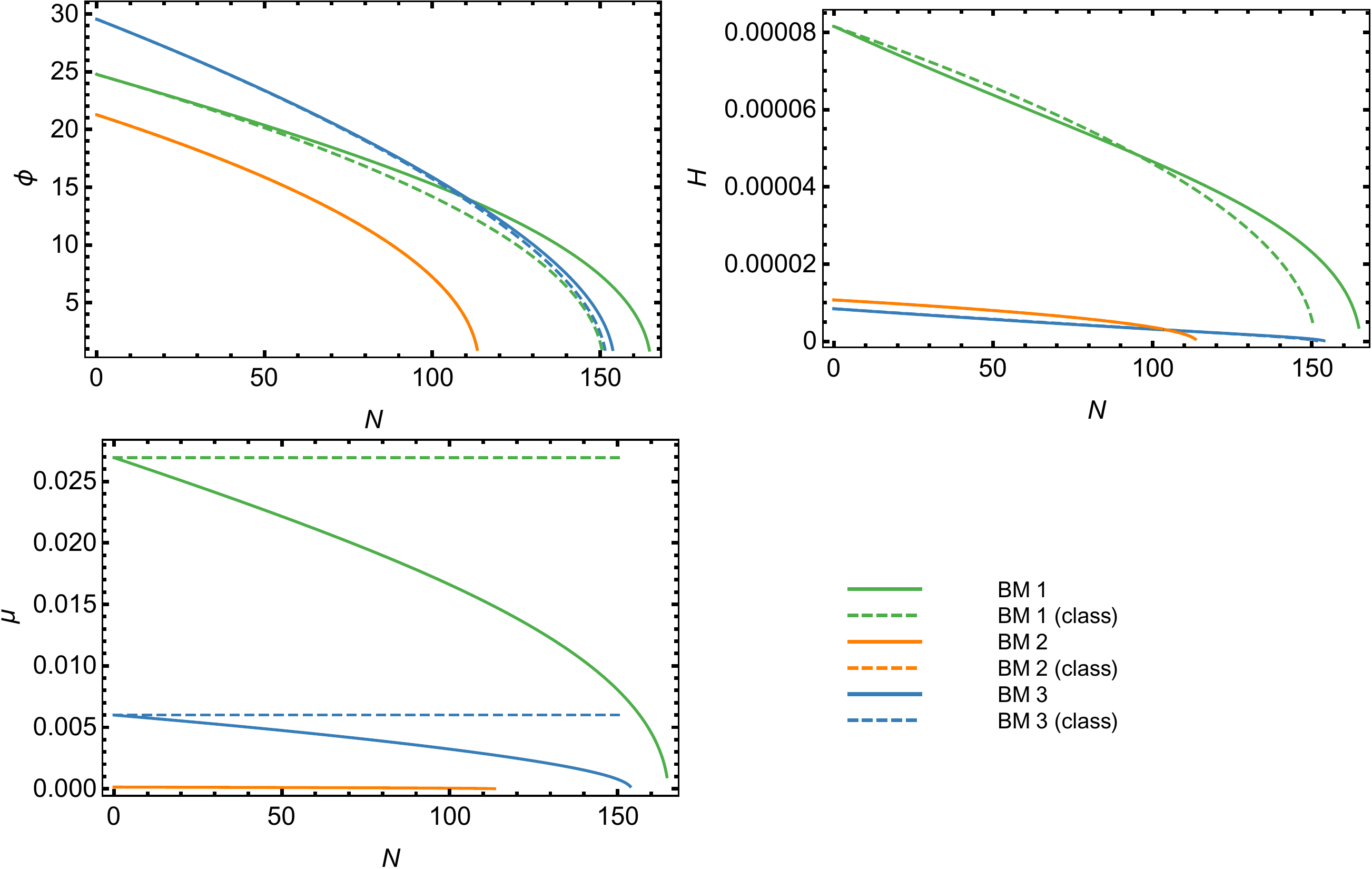}
\caption{Evolution of field value, Hubble rate and the inferred renormalisation scale. For large values of $g$ (BM 1) the RG-running of the parameters can have a noticeable effect on the cosmological evolution.}
\label{fig:TwoFieldsPhiHMuNBenchmark}
\end{center}
\end{figure*}

The relative change of $m_\phi^2$, $\lambda_\phi$, and $g$ due to RG-running is shown in Fig.~\ref{fig:TwoFieldsRGrunningBenchmark1}. For larger values of $g$ this change can be noticeable. They are reflected in the quantum evolution of the field shown in Fig.~\ref{fig:TwoFieldsPhiHMuNBenchmark}.

%%%%%%%%%%
\subsection{Full parameter scan}
\label{sec:scan2}
%%%%%%%%%%

It is tempting to search for parameter sets that maximize the effect of the quantum contributions. However, we want to convey an unbiased picture of their relevance. As before, we therefore sample the model parameters and initial conditions and evolve the non-slow roll equations of motion, computing the basic CMB observables. Marginalizing over $A_s$, we again generate a region in the $n_s-r$ plane.
To reduce the sampling parameter space somewhat, we take $\lambda_{\sigma,0}=0$, $m^2_{\sigma,0}=0$, $\xi_{\phi,0}=0$ and $\xi_{\sigma,0}=0$ throughout. Because they all run, they only vanish at $\phi_0$. This leaves as initial input $\lambda_{\phi,0}$, $m^2_{\phi,0}$ and the coupling $g_0$. We again take the ranges (\ref{eq:ranges}), and add
\begin{eqnarray}
g_0\in[10^{-16};10^{-4}].
\end{eqnarray}

\begin{figure}
\begin{center}
\includegraphics[width=8cm]{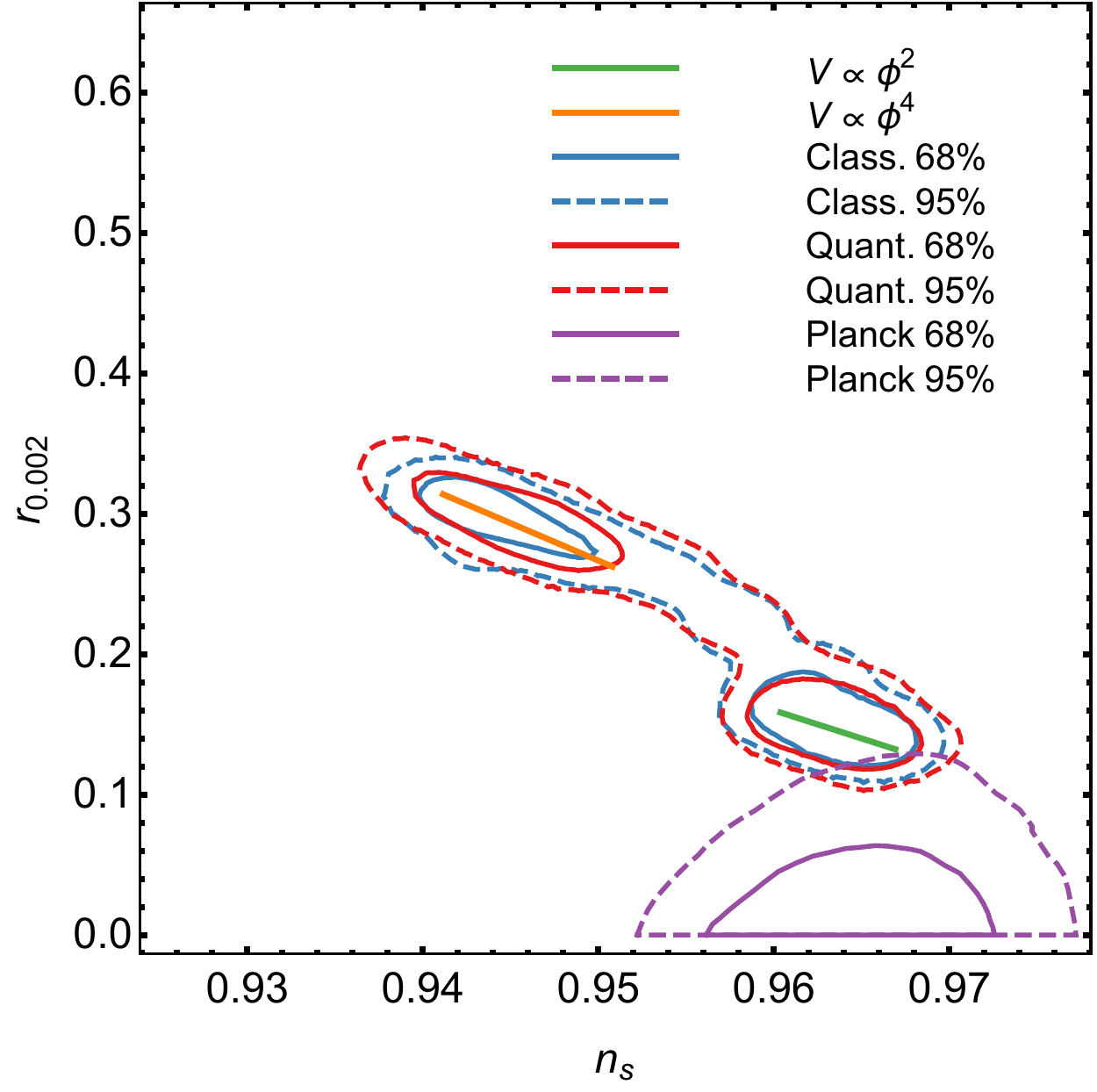}
\caption{The confidence regions for classical and quantum evolution the two-field case in the $n_s-r$ plane. Overlaid the Planck allowed range and the slow-roll monomial benchmarks (orange and green lines).}
\label{fig:quant2}
\end{center}
\end{figure}

Fig.~\ref{fig:quant2} shows the two-field classical and quantum regions consistent with observations in the $n_s-r$-plane. As before, the slow-roll monomial results are indicated by the orange and green line segments, and the Planck results are marked with purple lines. For the present sweep, $\xi_{\phi,0}$ is zero. As a consequence, the classical result is the "banana"-shaped $\phi^2+\phi^4$ region from Fig.~\ref{fig:1field}. Without the RG-running this agreement with the classical limit of the single-field model should be exact and, if $g$ is small, it should hold approximately, since the quantum contributions by the other non-vanishing parameters are tiny. 

We see that the overall differences between the evolution with both small $g$ and large $g$ are small. This is somewhat surprising, since we have seen that at least for trajectories of the BM1-type, both evolution and observables are shifted by some percent. Apparently, when sampling over the whole range, one recovers the same allowed region, even though individual trajectories (corresponding to a given parameter set) are moved around within this region. 

\begin{figure*}
\begin{center}
\includegraphics[width=\textwidth]{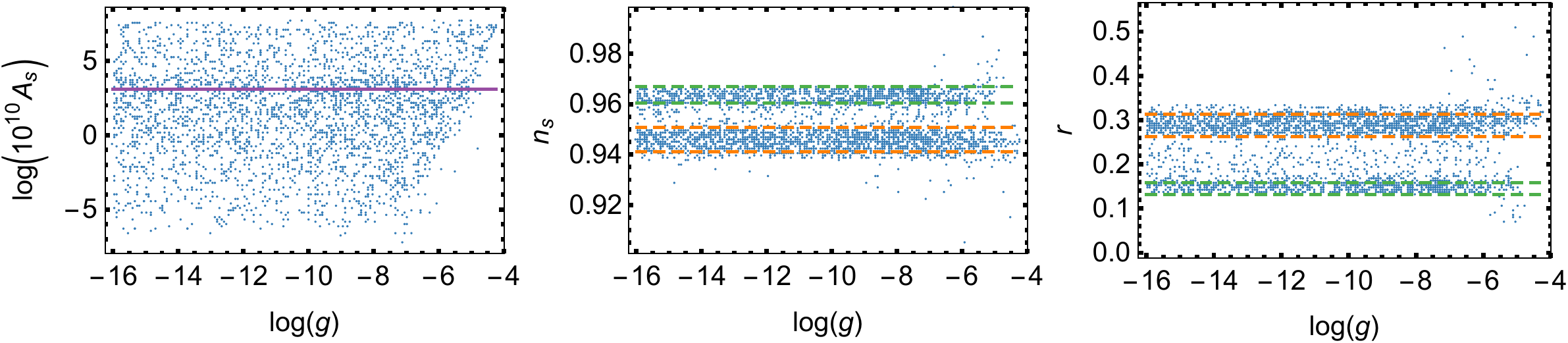}
\caption{Scatter plots showing the observables as a function of $g$. For large values we see a deviation from the region given by the slow-roll results for monomial inflation. However, for $g \gtrsim 10^{-5}$ the data fails to satisfy the experimental constraints on $A_s$ indicated by the solid purple horizontal line. For $\gtrsim10^{-4}$ inflationary evolution does not occur at all.}
\label{fig:TwoFieldAsGNsGRG}
\end{center}
\end{figure*}

One may speculate that even larger values of $g$ will lead to larger effects such that the $n_s -r$ contour is brought into overlap with the Planck allowed region. However, we find that for $g \gtrsim 10^{-5}$ the observed value for $A_s$ is not met any more (see Fig.~\ref{fig:TwoFieldAsGNsGRG}). And as $\gtrsim 10^{-4}$, inflation is ruined altogether by the running of $\lambda_\phi$ sourced by large $g$. It is possible that carefully tuning to the narrow region around $g\simeq 10^{-5}$ one may find some singular cases with large corrections to the observables, but by a flat-logarithmic sampling as we do here, such a region does not show up.

%%%%%%%%%%%%%%
\subsection{Validating the choice of $\mu$}
\label{sec:optimal2}
%%%%%%%%%%%%%%%

\begin{figure}
\begin{center}
\includegraphics[width=\columnwidth]{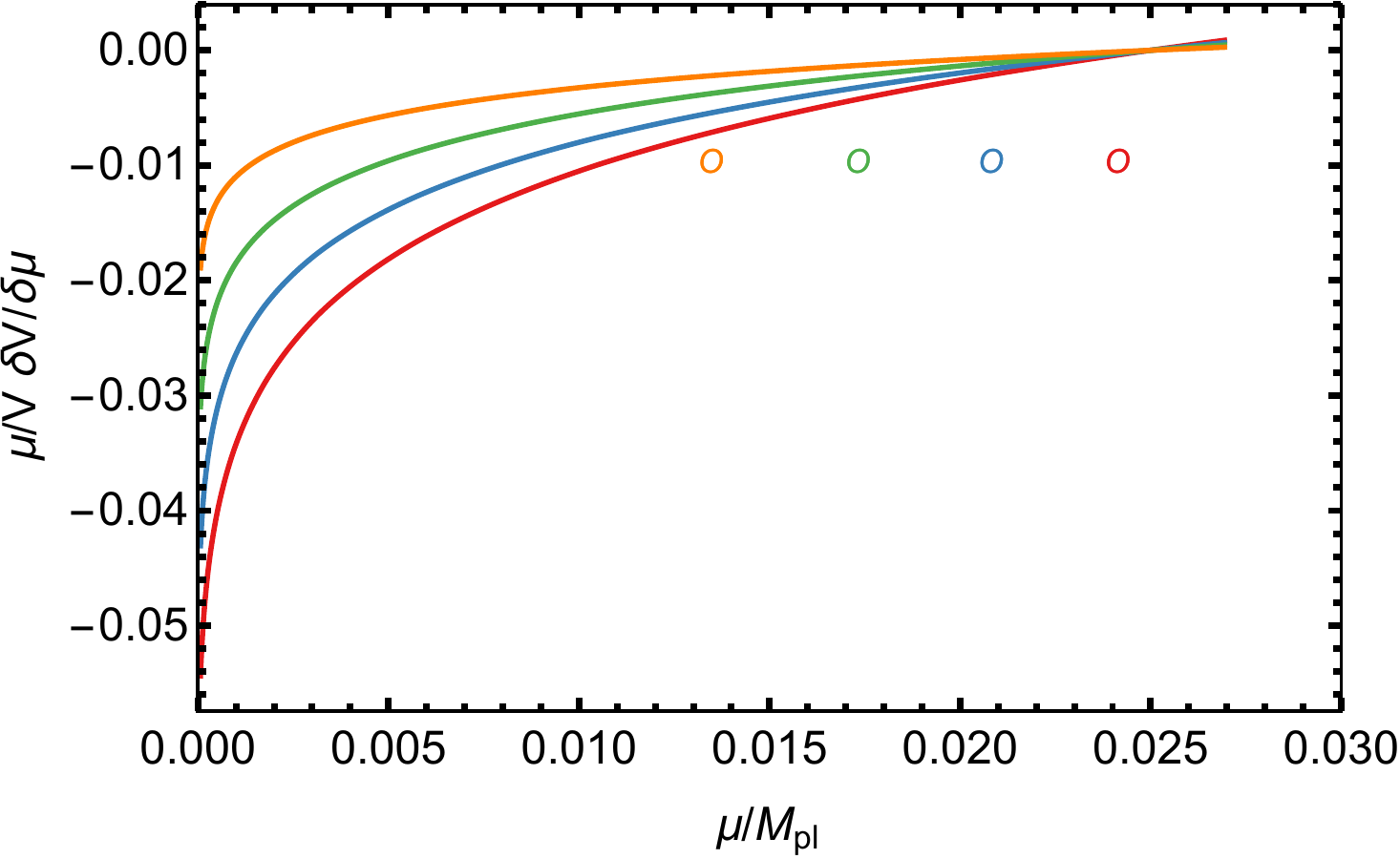}
\caption{The pseudo-optimal scale at different times (circles), and the potential as a function of $\mu$ given sets of $(\phi, H, \dot{H})$ along the trajectory.}
\label{fig:trajmu2}
\end{center}
\end{figure}
Along a trajectory in ($\phi$, $\dot{\phi}$, $H$, $\dot{H}$)-space, we again compute the relative dependence on the scale (\ref{eq:dVeff}). In Fig.~\ref{fig:trajmu2} is then the analogue of Fig.~\ref{fig:trajmu1}, but for two coupled fields. We see that the dependence on $\mu$ is much larger than for the 1-field model (Fig.~\ref{fig:trajmu1} was scaled up with $10^5$, Fig.~\ref{fig:trajmu2} is not), but still at the percent level for most of the range shown. Our time-dependent, pseudo-optimal choice of scale (circles) nicely traverse the range of scales, never coming near the singular region near the x-axis. In fact we see that as the pseudo-optimal scale decreases, the correspondingly colour-coded $\mu$ dependence of $V_{\rm eff}$ (red to orange) becomes more shallow. We have not observed a trajectory where the pseudo-optimal scale has caught up with the large $\mu$-dependence region. We conclude that our prescription for the choice of scale, apart from being very convenient also ensures small $\mu$-dependence.

%%%%%%%%%%%%%%
\section{Numerical implementation}
\label{sec:MMC}
%%%%%%%%%%%%%

In this section we provide some more details on how we solve the field- and Friedmann equations and how we sample the parameters.
Our numerical implementation is based on the public Fortran package MultiModeCode \cite{Price:2014xpa}. It is suited to the computation of observables for multi-field inflation models in the $\delta N$-formalism, or by solving the mode equations directly. In the present work we make use of the former feature only.

We however have to adapt the code significantly, in order to support the non-minimal coupling to gravity as well as the quantum corrected field (\ref{eq:FieldQuant}) and Friedmann equations (\ref{eq:FriedmannQuant}) (or equations (\ref{eq:TwoFieldQuant}) and (\ref{eq:TwoFriedmanQuant}) for the two-field model) that take a non-standard form.
They represent a coupled system of ordinary differential and algebraic (constraint) equations. 
This system determines the evolution of mean field values and the Hubble rate in time or, after a corresponding transformation, in the number of e-folds since the beginning of inflation. 
Usually, the algebraic first Friedman equation is used to explicitly solve for $H$.

With quantum corrections, however, its right-hand side depends in a complicated way on $H$. This is true even if the logarithms in (\ref{eq:FieldQuant}) and (\ref{eq:FriedmannQuant}) are eliminated by choice of the renormalisation scale, because $\mu$ then becomes a function of $H$ (as well as $\phi$ and $\dot{\phi}$). Recall that, in each step, $\mu$ is determined as the solution of the algebraic equation (\ref{eq:mu02}) for $\mu$.

To avoid the numerically solving the Friedmann equation, which would also need to be repeated in each time step of the evolution, we choose to solve the second Friedman equation for $dH/d N$. In the single-field case, this is possible explicitly if the quantum terms are made zero through the choice of $\mu$ described above. In the two-field case, the choice of $\mu$ that eliminates the direct quantum contributions to the mean field equation leaves a non-vanishing quantum contribution to both Friedmann equations. As argued above, for the purposes of the present paper we can neglect these terms in numerical computations. 

For both models, we then arrive at the first order system (with $m_\phi^2 = m^2$, $\lambda_\phi = \lambda$ and  $\xi_\phi = \xi$ in the 1-field case)
\begin{align}
\label{eq:HdotExplicit}
&\phi'=v\,,\\
&{v'}=\frac{H'v}{H}\nonumber\\&-\frac{1}{H^2}\bigg(3H^2 {v}+\left(m^2_\phi+\xi_\phi 6H(H'+2H)\right)\phi +\frac{1}{6}\lambda_\phi\phi^3\bigg) \,,\nonumber\\
&{H'}=\frac{H^{-1}}{48 \left(\xi_\phi  (6 \xi_\phi -1) \phi ^2+1\right)}\nonumber\\&\qquad\times\big[ 12 H^2 (6 \xi_\phi  (1-8 \xi_\phi ) \phi ^2\nonumber\\&\qquad\qquad-4 \xi_\phi  \phi  v+(4 \xi_\phi -1) \phi '^2-6)\nonumber\\
&\qquad\qquad +  \phi ^2 \left(\lambda_\phi  (1-8 \xi_\phi ) \phi ^2+m_\phi^2 (12-48 \xi_\phi )\right)\big]\,,\nonumber
\end{align}
where the primes denote the derivative with respect to the number of e-folds, $\phi' =d\phi/dN$ etc. 
The cost of this approach is that we need to solve the differential equation for $H$ alongside hose for the fields.\footnote{Combined with the fact that the field equation is no longer determined by the classical potential that also governs the potential slow-roll parameters this constitutes the main change that needs to be made to an inflation solver like MultiModeCode.}
All model parameters in (\ref{eq:HdotExplicit}) are functions of $\mu=\mu(H,\phi,\phi')$ which needs to be computed numerically in each step. This solution of (\ref{eq:mudef1}) or (\ref{eq:mu_2field}) determines the overall numerical costs.

The first Friedman equation is used, together with the field-equation, in slow-roll approximation to determine the initial conditions self-consistently:
\begin{align}
\label{eq:ICs}
&0=3H_0^2 {\phi_0'}\\& +\left(m^2_{\phi,0}+\xi_{\phi,0} 6H_0(H_0'+2H_0)\right)\phi_0+\frac{1}{6}\lambda_{\phi,0}\phi_0^3 \,,\nonumber\\
&3 H_0^2 =\frac{\frac12 H_0^2 \phi_0'^2 + \frac12 m_{\phi,0}^2\phi_0^2 + \frac{1}{24}\lambda_{\phi,0}\phi_0^4 + 6 H_0^2\phi_0'\phi_0 \xi_{\phi,0}}{1 - \phi_0^2 \xi_{\phi,0}}\,,\nonumber\\
&{H_0'}=\frac{H_0^{-1}}{48 \left(\xi_{\phi,0}  (6 \xi_{\phi,0} -1) \phi_0^2+1\right)}\nonumber\\&\qquad\times\big[ 12 H_0^2 (6 \xi_{\phi,0}  (1-8 \xi_{\phi,0} ) \phi_0^2\nonumber\\&\qquad\qquad -4 \xi_{\phi,0}  \phi_0  \phi_0'+(4 \xi_{\phi,0} -1) \phi_0'^2-6)\nonumber\\
&\qquad\qquad +  \phi_0^2 \left(\lambda_{\phi,0}  (1-8 \xi_{\phi,0} ) \phi_0^2+m_{\phi,0}^2 (12-48 \xi_{\phi,0} )\right)\big]\,.\nonumber
\end{align}
Given the initial model parameters and an initial value for the field, $\phi_0$, this non-linear algebraic system determines $H_0$, $\phi_0'$ and $H_0'$. It can be solved explicitly, but we solve it numerically for accuracy reasons.

The determination of consistent slow-roll initial conditions as well as that of the renormalisation scale, requires solving non-linear systems of equations.
We have linked the MultiModeCode package with Mathematica's WSTP (former MathLink library). This allows to reach the level of precision required to not pollute MultiModeCode's adaptive solving of the system of differential equations that govern inflation and to expose the typically small quantum effects. We can perform these computations at arbitrary precision and need not re-express the lengthy quantum contributions in numerically stable form for the many possible hierarchies of the parameters.

Throughout the paper we have chosen flat prior distributions for the logarithms of all model parameters at $\mu=\mu_0$, each on the intervals quoted in the sections. For the field $\phi$ we generate initial values on the interval $[2;30]$ obeying a flat non-logarithmic distribution. As the pivot scale $k_*$ we take the value $k_*=0.002$ everywhere. $N_*$ varies on the flat prior $[50;60]$ throughout. To determine the end of inflation we use the condition $\epsilon_1 = 1$. Sampled parameter sets for which not sufficiently many e-folds are achieved or for which inflation continues indefinitely are ignored.

Based on the obtained numerical values for the observables $A_s$, $n_s$ and $r$ we estimate their multi-variate distribution $P(A_s,n_s,r)$ and marginalize it with a Gaussian for $\log (10^{10} A_s)$, centred at the Planck experimental value, to obtain $P(n_s,r)$. The confidence regions in Fig.~\ref{fig:1field}, \ref{fig:1fieldnonmin} and \ref{fig:quant2} are given by the contour lines for the values $P_{\alpha}$ ($\alpha=95.45\%,\,68.27\%$) for which
\ba
\int_{-\infty}^{+\infty} \theta (P(n_s,r)-P_{\alpha}) P(n_s,r) d n_s d r = \alpha\,.
\ea
As explained, the parameters are constrained by the requirement for successful inflation. To perform a Bayesian parameter estimation a Monte-Carlo sampling method would need to be employed, given the high-dimensionality of the parameter space. In view of the moderate influence of the quantum contributions in the considered models, and the entailed additional numerical complexities, we have not attempted this in the present work, relying on information from scatter plots. Some of these are displayed above.

%%%%%%%%%%%%%%%%%%
\section{Conclusion}
\label{sec:conclusion}
%%%%%%%%%%%%%%%%%%%%%%%%%%%%%%

We have studied quantum corrections to inflation, and their impact in ruling certain models in and out of the Planck-allowed region in $n_s-r$ space. 
Specifically, we considered a single-field model with quadratic and quartic terms, coupled non-minimally to gravity as well as a two-field model in which the second field has a portal coupling with the inflaton field. We simplified the computations by taking the second field to be in its vacuum, mimicking spectator fields coupled to the Standard Model. The quantum corrections enter in the form of the RG-running of the model parameters and quantum corrections to the energy momentum tensor that contribute to field and Friedmann equations. 

The size of the latter depends on the choice of the renormalisation scale. We show that they can be made zero in the single-field and very small in the two-field case, leaving effectively only the RG-running to be considered. Although not the truly optimal choice of $\mu$, we argued that the precise value was not crucial. For the single-field model, the smallness of the inflaton self-coupling made the running negligible altogether. Whereas in the two-field model, the running was still sizeable. 

At the classical level, we confirmed that non-minimal couplings to gravity can bring the considered models into agreement with Planck measurements. In order to obtain an unbiased statement about the relevance of the quantum corrections caused by the RG-running of the parameters, we sampled the model parameters on broad intervals compatible with existing constraints. We found that the quantum corrections are negligible in the single-field case. 

In the second case with two fields, we found that the RG-running can influence the trajectories of the field notably if the portal coupling takes large enough values. However, the overall effect on the observables does hardly affect the contours of the posterior probability distribution for the observables. It seems that individual trajectories are swapped around inside the contours, without substantially altering them. Interestingly, we found that large values of the portal coupling tend to ruin inflation altogether putting a limit of $g_0<10^{-4}$, up to details of the scale $\mu_0 (\phi_0)$ where $g_0$ is introduced. This bound also restricts the size of the quantum corrections.

In conclusion, a correct and reliable implementation and analysis of quantum corrections to inflation involves considering RG-running, allowing for a time- and field-dependent RG-scale, including curvature corrections and imposing observational limits on $A_s$. We find that $\phi^2+\phi^4$-inflation coupled to a spectator field is as disfavoured at the quantum level as for the classical approximation. We did not do explicit scans for $\xi_0\neq 0$ for the two-field model, but suspect that a similar conclusion may apply. 

The obvious next steps include adapting our procedure to other models of inflation, as well as including IR effects arising from resummations of diagrams (see for instance \cite{Herranen:2013raa}). Of particular interest would be to consider relaxing the semi-classical approach to include both scalar and tensor metric degrees of freedom in our quantum treatment \cite{Herranen:2015aja}. This poses a number of other challenges to do with the running of gravitational couplings and renormalizability.

%%%%%%%%%%%%%

\acknowledgments{
MH and AH and AT are supported by the Villum Foundation Grant No.~YIP/VKR022599. }

%% bibliography

\end{document}